\documentstyle[aps,pra,eqsecnum,graphics,epsf]{revtex}

\begin{document}
\title{Decoherence-Free Subspaces for Multiple-Qubit Errors: (II)
Universal, Fault-Tolerant Quantum Computation}
\author{Daniel A. Lidar,$^{1}$ Dave Bacon,$^{1,2}$ Julia
Kempe$^{1,3,4}$ and K.B. Whaley$^{1}$}
\address{Departments of Chemistry$^{1}$, Physics$^{2}$ and\\
Mathematics$^{3}$,\\
University of California, Berkeley, CA 94720\\
\'{E}cole Nationale Superieure des T\'{e}l\'{e}communications,\\
Paris, France$^4$}
\date{\today}
\maketitle

\begin{abstract}
Decoherence-free subspaces (DFSs) shield quantum information from errors
induced by the interaction with an uncontrollable environment. Here we study
a model of correlated errors forming an Abelian subgroup (stabilizer) of the
Pauli group (the group of tensor products of Pauli matrices). Unlike
previous studies of DFSs, this type of errors does not involve any spatial
symmetry assumptions on the system-environment interaction. We solve the
problem of universal, fault-tolerant quantum computation on the associated
class of DFSs.

\end{abstract}

\section{Introduction}

Methods to protect fragile quantum superpositions are of paramount
importance in the quest to construct devices that can reliably process
quantum information \cite{Lo:book,Williams:book98}. Compared to their
classical counterparts such devices feature spectacular advantages in both
computation and communication, as discussed in a number of recent reviews 
\cite{Steane:98,Aharonov:98,Cleve:99}. The dominant source of the fragility
of a quantum information processor (QIP) is the inevitable interaction with
its environment. This coupling leads to {\em decoherence}: a process
whereby coherence of the QIP-wavefunction is gradually destroyed. Formally,
the evolution of an open system (coupled to an environment)\ such as a QIP
can be described by a completely-positive map \cite{Alicki:87}, which can
always be written in the explicit form known as the Kraus operator sum
representation \cite{Kraus:83}: 
\begin{equation}
\rho (t)=\sum_{d}A_{d}(t)\rho (0)A_{d}^{\dagger }(t).
\end{equation}
Here $\rho $ is the system density matrix, and the ``Kraus operators'' $
\{A_{d}\}$ are time-dependent operators acting on the system Hilbert space,
constrained only to sum to the identity operator: $\sum_{d}A_{d}^{\dagger
}A_{d}=I$ (to preserve ${\rm Tr}[\rho ]$).\footnote{
As shown, e.g., in \cite{Bacon:99}, the operator sum representation can be
derived from a Hamiltonian model by considering the reduced dynamics of a
system coupled to a bath $B$: $\rho (t)={\rm Tr}_{B}[U(t)(\rho (0)\otimes
\rho _{B}(0))U^{\dagger }(t)]$. Here the trace is over the bath degrees of
freedom, $U=\exp (-iH_{SB}t)$ is the unitary evolution operator of the
combined system-bath, and $H_{SB}$ is their interaction Hamiltonian. One
finds: $A_{d=(\mu ,\nu )}=\sqrt{\mu }\langle \mu |U|\nu \rangle $, where $
|\mu \rangle ,|\nu \rangle $ are bath states in the spectral decomposition
of the bath density matrix:$\;\rho _{B}=\sum \mu |\mu \rangle \langle \mu |$.} Decoherence is the situation where there are at least two Kraus operators
that are inequivalent under scalar multiplication. The Kraus operators are
in that case related to the different ways in which errors can afflict the
quantum information contained in $\rho $ \cite{Nielsen:98}. Conversely, if
there is only one Kraus operator, then from the normalization condition it
must be unitary: $A=\exp (-iHt)$ with $H$ Hermitian, so that $\rho $
satisfies the {\em closed}-system Liouville equation $\dot{\rho}=-i[H,\rho ]$, $H$ being the system Hamiltonian. In this case there is no decoherence.

Two principal {\em encoding} methods have been proposed to solve the
decoherence problem: (i) Quantum Error Correcting Codes (QECCs) \cite
{Shor:95,Calderbank:96,Steane:96a,Bennett:96a,Kitaev:96,Gottesman:97,Knill:97b}
(for a recent review see \cite{Steane:99}), (ii) Decoherence-Free Subspaces (DFSs) \cite
{Zanardi:97c,Zanardi:97a,Duan:98,Duan:98a,Lidar:PRL98,Lidar:PRL99,Knill:99a}
, also known as ``noiseless'', or ``error-avoiding quantum codes''. In both
methods quantum information is protected against decoherence by encoding it
into ``codewords'' (entangled superpositions of multiple-qubit states) with
special symmetry properties. To exhibit these, it is useful to expand the
Kraus operators over a fixed operator basis. For qubits a particularly
useful basis is formed by the elements of the Pauli group:\ the group of
tensor products of all Pauli matrices $\{\sigma _{k}^{\alpha _{k}}\}$, where 
$\alpha =0,x,y,z$ ($\sigma ^{0}$ is the $2\times 2$ identity matrix) and $
k=1..K$ is the qubit index. An element of the Pauli group can be written as $
E_{a}=\otimes _{k=1}^{K}\sigma _{k}^{\alpha _{k}}$, where $a=(\alpha
_{1},...,\alpha _{K})$. The $4^{K+1}$elements $\{E_{a}\}$ of the Pauli group
(we include factors of $\pm $,$\pm i$ in this count) square to identity, are
both unitary and Hermitian, either commute or anti-commute, and satisfy $
{\rm Tr}[E_{a}^{\dagger }E_{b}]=\delta _{ab}/2^{K}$. When the Kraus
operators are expanded as 
\begin{equation}
A_{d}(t)=\sum c_{ad}(t)E_{a},
\end{equation}
the operators $\{E_{a}\}$ acquire the significance of representing the
different physical errors that can corrupt the quantum information. The
weight $w(E_{a})$ is the number of non-zero $\alpha _{k}$ in $a$. Let
us now assume a short-time expansion of the $c_{ad}(t)$ (relative to
the bath-correlation time). The
situation where only those $E_{a}$ with $w(E_{a})=1$ have non-vanishing $
c_{ad}(t)$ is called the ``independent errors'' model (assuming the
$c_{ad}$, which are essentially bath correlation functions
\cite{Bacon:99}, are 
statistically independent). Correlated errors corresponds to the situation
where some $E_{a}$ with $w(E_{a})>1$ have non-vanishing $c_{ad}(t)$:\ two or
more qubits are acted upon non-trivially with the same coefficient $c_{ad}$.
QECCs can be classified according to the maximum weight of the errors they
can still correct (this is related to the notion of a ``distance''\ of a
code \cite{Steane:99}). QECCs can generally deal at least with errors of
weight $1$. Barring accidental degeneracies, non-trivial DFSs, on the other
hand, generally do not exist if there are errors with weight $1$ \cite
{Lidar:PRL98}. To make these ideas more precise, let us briefly recall the
definitions of QECCs and DFSs.

A QECC is a subspace ${\cal C}={\rm Span}[\{|i\rangle \}]$ of the system
Hilbert space with the symmetry property that different errors take
orthogonal codewords $|i\rangle $ and $|j\rangle $ to orthogonal states \cite
{Knill:97b}:

\begin{equation}
\langle i|E_{a}^{\dagger }E_{b}|j\rangle =\gamma _{ab}\delta _{ij}.
\label{eq:QECC-cond}
\end{equation}
Here $\gamma _{ab}$ are the elements of an Hermitian matrix $\gamma $ and $
\delta _{ij}$ is the Kronecker delta. This property ensures that if an error 
$E_{a}$ occurs it can be detected and subsequently reversed \cite{Knill:97b}
. A large variety of QECCs have been found \cite{Steane:99}. A particularly
useful and large class, one which will occupy our attention in this paper,
arises when one considers Abelian subgroups $Q$ of the Pauli group. Given
such an Abelian Pauli-subgroup, or {\em stabilizer} $Q$ (we will use both
terms interchangeably in this paper), its $+1$ eigenspace is a QECC known as
a {\em stabilizer code} \cite{Gottesman:97}. The set of errors $\{E_{a}\}$
is correctable by this code if for every two errors $E_{a},E_{b}$ there
exists some $q\in Q$ such that 
\begin{equation}
\{E_{a}^{\dagger }E_{b},q\}=0.  \label{eq:QECC-stab}
\end{equation}
This is because under the stipulated condition $\langle i|E_{a}^{\dagger
}E_{b}|j\rangle =\langle i|E_{a}^{\dagger }E_{b}q|j\rangle =-\langle
i|qE_{a}^{\dagger }E_{b}|j\rangle =-\langle i|E_{a}^{\dagger }E_{b}|j\rangle 
$ so that $\langle i|E_{a}^{\dagger }E_{b}|j\rangle \propto \delta _{ij}$ 
\cite{Gottesman:97}: the QECC condition [Eq.~(\ref{eq:QECC-cond})] is
satisfied. To correct an error $E_{a}$ one simply applies the unitary
operator $E_{a}^{\dagger }$ to the code. Note that this involves active
intervention:\ measurements to diagnose the error, and error reversal.

DFSs can be viewed as highly ``degenerate'' QECCs, where degeneracy refers
to the rank of $\gamma $:\ DFSs are rank-1 QECCs (i.e., $\gamma _{ab}=\gamma
_{a}\gamma _{b}$) \cite{Lidar:PRL99,Duan:98d}. Equivalently, a DFS can be
defined as the simultaneous eigenspace ${\tilde{{\cal H}}}$ $={\rm Span}[\{|
\tilde{j}\rangle \}]$ of all Kraus operators \cite{Lidar:PRL99}: 
\begin{equation}
A_{d}|\tilde{j}\rangle =a_{d}|\tilde{j}\rangle
\end{equation}
($\{a_{d}\}$ are the eigenvalues). Viewed in this way, DFSs have the
remarkable property that they offer complete protection for quantum
information without the need for any active intervention:\ $\tilde{\rho}
(t)=\sum_{d}A_{d}(t)\tilde{\rho}(0)A_{d}^{\dagger }(t)=\tilde{\rho}
(0)\sum_{d}|a_{d}|^{2}=\tilde{\rho}(0)$, for $\tilde{\rho}$ with support
exclusively on ${\tilde{{\cal H}}}$. Thus a DFS is a ``quiet corner''\ of
the system Hilbert space, which is completely immune to decoherence. Like
stabilizer-QECCs, DFSs can also be characterized as the $+1$ eigenspace of a
stabilizer, which however is generally {\em non-Abelian} over the
Pauli group \cite
{Bacon:99a,Kempe:00} (i.e., a DFS is generally a non-additive code
\cite{Rains:97}). Most work on DFSs to date has focused on a model of 
highly correlated errors, known as ``collective decoherence''. In this model
the (non-Abelian) stabilizer is composed of tensor products of identical $
SU(2)$ rotations + contractions on all qubits. Here we will not concern
ourselves with the collective decoherence model, and the term stabilizer
will be reserved for the Abelian subgroups of the Pauli group.

In a companion paper \cite{Lidar:00a} (referred to from here on as
``paper 1'') we began a study of DFSs for non-collective errors. We derived
a necessary and sufficient condition for a subspace to be decoherence-free
when the Kraus operators are expanded as linear combinations over the
elements of an arbitrary group. The decoherence-free states were shown to be
those states that transform according to the one-dimensional irreducible
representations (irreps) of this group. As above, it is natural to focus on
the case where this group is the Pauli group. This is so not only because of
the connection to stabilizer-QECCs, but also because the Pauli group arises
in the context of many-qubit systems, where it is often natural to expand
the Hamiltonians in terms of tensor products of Pauli matrices. To find
DFSs, therefore, we focus here on subgroups of the Pauli group. Note that
the non-Abelian subgroups of the Pauli group do not have one-dimensional
irreps \cite{Lidar:00a}, and hence in this case a DFS can be associated
only with the Abelian subgroups (which of course have only one-dimensional
irreducible representations).

We can now define the error model that will concern us in this paper. Unlike
the stabilizer-QECCs case, where the errors that the code can correct are
those that anti-commute with the stabilizer, {\em in the DFS case the errors
are the elements of the stabilizer itself}. We shall refer to these errors
as ``stabilizer-errors''. The Abelian subgroups of the Pauli group cannot
contain single-qubit operators, since these would generally generate the
whole Pauli group.\footnote{The exceptions are: (i) The subgroup
operators have constant $\alpha
=x,y $ or $z$ -- the Pauli matrix index; (ii) The single-qubit
operators act only on those qubits where all other operators act as identity.} Hence as errors the elements of the subgroup represent 
{\em multiple-qubit} couplings to the bath. As explained above, this is
therefore a correlated-errors model, which is distinguished from previous
work on DFSs in that it does not involve any spatial-symmetry assumptions.
The physical relevance of this error-model was discussed in paper 1, and
will be embellished here. The DFS is not affected by these
stabilizer-errors, but the rest of the Hilbert space is and may decohere
under their influence. Several examples of DFSs corresponding to Abelian
subgroups were given in paper 1. Our purpose in this sequel paper is to
complete our study of this class of DFSs by showing how to perform universal
fault-tolerant quantum computation on them.

The central challenge in demonstrating universal fault-tolerant quantum
computation on DFSs is to show how this can be done using only 1- and 2-body
Hamiltonians, and a\ small number of measurements.\footnote{
By ``small'' we mean that the measurements do not have to be fast compared
to the bath correlation time. If they are then the decoherence is avoided
essentially by use of the quantum Zeno effect.} Several previous
publications have addressed the issue of universal quantum computation on
DFSs, but left this challenge unanswered \cite
{Lidar:PRL98,Zanardi:99a,Beige:00}. In Refs. \cite{Bacon:99a,Kempe:00} we
accomplished this task for the first time, in the collective decoherence
model. Collective decoherence is the situation where all qubits are coupled
in an identical manner to the bath, i.e., there is a strong {\em spatial
symmetry}: qubit permutation-invariance. In this case, by using exchange
operations, it is possible to implement universal quantum computation
without ever leaving the DFS. The procedure is therefore naturally
fault-tolerant. In the present paper we will show how to implement universal
fault-tolerant quantum computation on DFSs that arise from the Pauli
subgroup error model, without requiring any spatial symmetry assumption.
However, it will not be possible to do so without leaving the DFS, thus
exposing the states to the subgroup errors. As will be shown here,
fault-tolerance is obtained by using the encoded states twice: in a dual
DFS-QECC mode. This duality arises from the fact that the DFS remains a
perfectly valid QECC for the errors that the stabilizer anticommutes with.

There are several ways to achieve universal fault-tolerant quantum
computation on stabilizer-QECCs; e.g., use of the sets of gates \{Hadamard, $
\sigma _{z}^{1/2}$, Toffoli\}\cite{Shor:96,Kitaev:96}, or \{Hadamard, $
\sigma _{z}^{1/4}$, controlled-{\sc NOT}\} \cite{Boykin:99}. Additional
methods were provided in \cite{Gottesman:97a,Knill:97a}. Our construction
reverts to the early ideas on the implementation of universal quantum
computing:\ we use single-qubit $SU(2)$ operations and a controlled-{\sc NOT}
gate \cite{Deutsch:95,DiVincenzo:95,Sleator:95,Lloyd:95}, except that these
are {\em encoded} operations, acting on codewords (not on physical qubits).
In general such encoded operations involve multiple qubits, and are not
naturally available. The key to our construction is a method to generate
many-qubit Hamiltonians by composing operations on (at most) pairs of
physical qubits. This is done by selectively turning certain interactions on
and off. A difficulty is that the very first such step can transform the
encoded states and take them outside of the DFS. However, by carefully
choosing the interactions we turn on/off and their order, we show that the
transformed states become a QECC with respect to the stabilizer-errors that
the DFS was immune to. This fact is responsible for the fault-tolerance of
our procedure. After the final interaction is turned off, the states return
to the DFS, and are once again immune to the stabilizer-errors.

The structure of the paper is as follows. In Section \ref{Connection} we
briefly review the main result of paper 1, and the connection between the
DFSs considered here and stabilizer-QECCs. We then discuss in Section \ref
{FTmeaning} the meaning of fault-tolerance in light of the error-model
considered in this paper. In the following two sections we present the main new
ideas and results of this paper: in \ref{make-SU(2)} we show how to generate many-qubit
Hamiltonians by composing two- and single-qubit Hamiltonians, and in
\ref{SU(2)} prove the
fault-tolerance of this procedure. We use it to generate encoded $SU(2)$
operations on the DFS qubits. Section \ref{CNOT} shows how, by using similar
methods, we can fault-tolerantly perform encoded {\sc CNOT} operations on
the encoded qubits, thus coupling blocks of qubits and completing the set of
operations needed for universal computation. The final ingredient is
presented in Section \ref{Measurements}, where we show how to
fault-tolerantly measure the error syndrome throughout our gate
construction. While our main motivation in this paper is to study
computation on DFSs in the presence of stabilizer-errors, it is also
interesting to consider the implications of the techniques we develop here
to the usual model of errors that anticommute with the stabilizer. We
consider this question briefly in Section \ref{Outlook}, and show that our
methods provide a new way to implement universal quantum computation which
is fault tolerant with respect to error {\em detection}. We conclude and
summarize in section \ref{Summary}.

\section{Connection between Pauli Subgroup DFS\lowercase{s} and
Stabilizer Codes}

\label{Connection}

In paper 1 we proved the following result:

{\it Theorem 1.} --- Suppose that the Kraus operators belong to the group
algebra of some group ${\cal G}=\{G_{n}\}$, i.e., ${\bf A}
_{d}=\sum_{n=1}^{N}a_{d,n}G_{n}$. If a set of states $\{|\tilde{j}\rangle \}$
belong to a given {\em one-dimensional } irrep of ${\cal G}$, then the DFS
condition ${\bf A}_{d}|\tilde{j}\rangle =c_{d}|\tilde{j}\rangle $ holds. If
no assumptions are made on the bath coefficients $\{a_{d,n}\}$, then the DFS
condition ${\bf A}_{d}|\tilde{j}\rangle =c_{d}|\tilde{j}\rangle $ implies
that $|\tilde{j}\rangle $ belongs to a {\em one-dimensional} irrep of
${\cal G}$. 

This theorem provides a characterization of DFSs in terms of the
group-representation properties of the basis set used to expand the Kraus
operators in. There are good physical reasons to choose the Pauli group as
this basis set: as argued in paper 1, the Pauli group naturally appears as a
basis in Hamiltonians involving qubits. Furthermore, using the Pauli group
allows us to make a connection to the theory of stabilizer-QECCs. To see
this consider the identity irrep, for which each element $G_{n}$ in the
group ${\cal G}$ acts on a decoherence free state $|\psi \rangle $ as

\begin{equation}
G_{n}|\psi \rangle =|\psi \rangle .
\end{equation}
Choosing ${\cal G}$ from now on as a Pauli subgroup $Q$, the DFS fixed by
the identity irrep is a stabilizer code, where $Q$ is the stabilizer group.
As mentioned above, a stabilizer code is defined as the $+1$ eigenspace of
the Abelian group $Q$.\footnote{
The DFSs corresponding to the other 1D irreps can also be turned into
stabilizer codes by a redefinition of the subgroup, taking into account the
minus signs appearing in the irrep under question. This kind of freedom is
well known in the stabilizer theory of QECCs \cite{Gottesman:97}.} It is
thus clear that the states fixed by $Q$ play a dual role: {\em they are at
once a DFS with respect to the stabilizer errors and a QECC with respect to
the errors that anticommute with some element of} $Q$.

It is simple to verify that basic properties of stabilizer codes hold, e.g.,
that if the stabilizer group has $K-l$ generators, then the code space (in
this case the DFS) has dimension $2^{l}$ (i.e., there are $l$ encoded
qubits) \cite{Gottesman:97}. Indeed, the dimension of an Abelian group with $
K-l$ generators is $N=2^{K-l}$, and we showed in paper 1 that the dimension
of the DFS is $2^{K}/N=2^{l}$.

\section{The Meaning of Fault-Tolerance}

\label{FTmeaning}

The observation that the Pauli subgroup DFSs are stabilizer codes allows us
to employ some results from stabilizer theory, and aids in the analysis of
when it is possible to perform universal fault-tolerant computation on these
DFSs.\footnote{
The reader may wonder whether it should not be possible to simply take over
the results about universal fault tolerant computation from stabilizer
theory, and apply them directly in the present case. However, a problem is
encountered when that construction is applied to the error-model considered
here, because multiple-qubit errors may propagate back as (non-perturbative)
single-qubit errors due to interaction with a ``bare'' (non-DFS) ancilla.
We are indebted to Dr. Daniel Gottesman for pointing out this problem to us.}
Before delving into the analysis, however, we should clarify what we mean by
fault-tolerance in the present context. The usual meaning of
fault-tolerance, as it is used in the theory of QECC, is the following: an
operation (gate $U$) is {\em not} fault-tolerant if an error $E$ that the
code could fix before application of the gate has become an unfixable error ($UEU^{\dagger }$) after application of the gate. For example, a single qubit
phase error ($I\otimes Z$) becomes a two-qubit phase error ($Z\otimes Z$)
due to the application of a {\sc CNOT} gate \cite{Gottesman:97a}; if the
code used could only correct single-qubit errors then as a result of the 
{\sc CNOT} gate (unless it is applied transversally, i.e., not coupling
physical qubits involved in representing the same encoded qubit) this code
can no longer offer protection. In this scenario, therefore, the {\sc CNOT}
gate was not a fault-tolerant operation. Conversely, an operation {\em is}
fault-tolerant if the code offers the same protection against the errors
that appear after application of the operation ($UEU^{\dagger }$) as it does
against the errors before the operation ($E$).

A complementary (``Heisenberg'' \cite{Gottesman:98}) picture to the
(``Schr\"{o}dinger'') description above is to consider the errors as
unchanged and the code ${\cal C}$, as well as the stabilizer $Q$, as
transformed after the application of each gate:\ ${\cal C}\longmapsto U{\cal 
C}$ and $Q\longmapsto UQU^{\dagger }$. Then fault-tolerance can be viewed as
the requirement that the new code is capable of correcting the original
errors. This point of view will be particularly useful for our purposes. In
our case the original errors are the elements of the Pauli subgroup $Q$ (the
stabilizer), and the gates $U$ will turn out not to preserve the original
code. Nevertheless, we will show that to the new stabilizer $Q^{\prime
}=UQU^{\dagger }$ corresponds a QECC (the transformed code ${\cal C}^{\prime
}=U{\cal C}$) that can correct the original errors. In this way the
fault-tolerance criterion is satisfied.

\section{Encoded $SU(2)$ from Hamiltonians}
\label{make-SU(2)}

We now begin in earnest our discussion of how to implement universal,
fault-tolerant quantum computation on the Pauli-subgroup DFSs. In this
section we show how arbitrary single encoded-qubit operations can be
implemented fault-tolerantly. We will do so by generating the entire encoded 
$SU(2)$ group from at most two-qubit Hamiltonians. We assume that the system
Hamiltonian is of the general two-qubit form 
\begin{equation}
H_{S}=\sum_{i=1}^{K}\sum_{\alpha =\{x,y,z\}}\omega _{i}^{\alpha _{i}}\sigma
_{i}^{\alpha _{i}}+\sum_{i>j=1}^{K}\sum_{\alpha ,\beta
=\{x,y,z\}}J_{ij}^{\alpha _{i}\beta _{j}}\sigma _{i}^{\alpha _{i}}\otimes
\sigma _{j}^{\beta _{j}},  \label{eq:Hs}
\end{equation}
with controllable parameters $\{\omega _{i}^{\alpha _{i}}\}$, $
\{J_{ij}^{\alpha _{i}\beta _{j}}\}$.

\subsection{Background}

Suppose we are given an error subgroup $Q$ generated by the elements $
\{q_{i}\}_{i=1}^{|Q|}$. By the results of paper 1 we know how to identify
the corresponding DFS, which is also a stabilizer-code with respect to the
errors that anti-commute with $Q$. This QECC aspect will not be needed as
long as we are only interested in {\em storing} information in this DFS:
then the $Q$-errors will have no effect. However, here we are interested in
the more ambitious goal of {\em computing} in the presence of the $Q$
-errors, which means that we must be able to implement logic gates. As
discussed above, these gates will take the states out the DFS and expose
them to the $Q$-errors. To be able to compute we will need some basic
results from the theory of fault-tolerant quantum computation using
stabilizer codes, as developed primarily in Ref.~ \cite{Gottesman:97a}. Let
us briefly review these results.

The set of operators which commute with the stabilizer themselves form a
group called the {\em normalizer} of the code, $N(Q)$. These elements are of
interest because they are operations which preserve the DFS. Let $q\in Q$, $
|\psi \rangle \in $ DFS$(Q)$; if $n\in N(Q)$ then 
\begin{equation}
q\left( n|\psi \rangle \right) =nq|\psi \rangle =n|\psi \rangle ,
\end{equation}
so that $n|\psi \rangle $ is in the DFS as well. Clearly, the stabilizer $Q$
is in the normalizer $N(Q)$ and so the only operations which act
nontrivially on the subspace are those which are in the normalizer but not
in the stabilizer: $N(Q)/Q$. While this means that these operations can be
used to perform useful manipulations on the DFS, it also means that if they
act uncontrollably, then they appear as errors that the code {\em cannot}
detect. As will be seen later on, these are both crucial aspects in our
construction.

For any Pauli-subgroup stabilizer code, the normalizer is generated by the
single qubit $\overline{X}_{i}$ and $\overline{Z}_{i}$ operations, where $
i=1,...,l $ labels the {\em encoded} qubits \cite{Gottesman:97a}. The
bar superscript denotes that these are ``encoded operations'': they perform
a bit-flip and a phase-flip on the encoded qubits. The gates $\overline{X}_{i}$
and $\overline{Z}_{i}$, however, are by themselves insufficient for universal
quantum computation. The usual stabilizer-QECC construction deals with
(typically {\em un}correlated) errors that anticommute with the stabilizer.
In this case, in addition to generating the normalizer of the Pauli group $
N(P_{K})$, one other operation is needed, such as the Toffoli gate \cite
{Shor:96}. Such constructions have been covered in several recent
publications \cite{Shor:96,Boykin:99,Gottesman:97a,Knill:97a,Calderbank:98}.
However, as emphasized above, the errors here are qualitatively different:
not only are they always correlated, rather than anticommuting with the
stabilizer, {\em the errors are the stabilizer itself}. Thus the usual
construction does not apply, and we introduce a different approach. We show
how to perform universal fault-tolerant quantum computation using the
early $SU(2)+${\sc CNOT} construction \cite
{DiVincenzo:95,Lloyd:95,Barenco:95a}, but applied to encoded (DFS) qubits.

\subsection{A Useful Formula: Conjugation by $\protect\pi /4$}

Instead of treating $\overline{X}$ and $\overline{Z}$ as gates, as in the usual
stabilizer-QECC construction, we employ them as {\em Hamiltonians}. Since $
\overline{X}$ and $\overline{Z}$ are in the normalizer, so are $\exp (i\theta \overline{X})$
and $\exp (i\theta \overline{Z})$, and so are any other encoded $SU(2)$
group (denoted $
\overline{SU(2)}$) operations obtained from them. By applying operations
from $\overline{SU(2)}$ alone we ensure that the code is preserved. To
obtain other $\overline{SU(2)}$ operations from $\overline{X}$ and $\overline{Z}$, we
use the Euler angle construction \cite{Rose:book}, which shows
that any rotation can be composed out of rotations about only two orthogonal
axes: 
\begin{equation}
\exp [-i\omega ({\bf n}\cdot {\bf \sigma })/2]=\exp (-i\beta \sigma
_{z}/2)\exp (-i\theta \sigma _{y}/2)\exp (-i\alpha \sigma _{z}/2).
\label{eq:Euler}
\end{equation}
Here the resulting rotation is by an angle $\omega $ about the direction
specified by the unit vector ${\bf n}$, both of which are functions of $
\alpha $,$\beta$, and $\theta $. Using Eq.~(\ref{eq:Euler}) and the mapping $
\{\sigma _{x},\sigma _{y},\sigma _{z}\}\longmapsto \{\overline{X},\overline{Y},\overline{Z}
\}$, we can construct any element of $\overline{SU(2)}$. To do so, we now
derive a form of the Euler angle construction which is particularly relevant
to operations with Pauli matrices. Assume that $A$ and $B$ are both tensor
products of Pauli matrices (and thus square to identity). Then: 
\begin{eqnarray}
\exp (-i\varphi A)B\exp (+i\varphi A) &=&(I\cos \varphi -Ai\sin \varphi
)B(I\cos \varphi +Ai\sin \varphi )  \nonumber \\
&=&B\cos ^{2}\varphi +ABA\sin ^{2}\varphi -i\sin \varphi \cos \varphi
\lbrack A,B]  \nonumber \\
&=&\left\{ 
\begin{array}{r}
B\quad {\rm if\,}[A,B]=0 \\ 
B\cos 2\varphi +iBA\sin 2\varphi \quad {\rm if\,}\{A,B\}=0
\end{array}
\right.  \label{eq:result1}
\end{eqnarray}
For the special case of $\varphi ={\pi /4}$ we define the conjugation with $
A $ by 
\begin{eqnarray}
T_{A}\circ \exp (i\theta B) &\equiv &\exp (-i\frac{\pi }{4}A)\exp (i\theta
B)\exp (+i\frac{\pi }{4}A)  \nonumber \\
&=&\left\{ 
\begin{array}{r}
\exp (i\theta B)\quad {\rm if\,}[A,B]=0 \\ 
\exp [i\theta (iAB)]\quad {\rm if\,}\{A,B\}=0
\end{array}
\right. .  \label{eq:result2}
\end{eqnarray}
This can be understood geometrically as a rotation by $\varphi =\pi /4$
about the ``axis'' $A$, followed by a rotation by $\theta $ about $B$,
followed finally by a $\beta =-\pi /4$ rotation about $A$, resulting overall
in rotation by $\theta $ about the ``axis'' $iAB$. All $\varphi =\pi /4$
rotations about a Pauli group member are elements of the normalizer of the
Pauli group: they take elements in the Pauli group under conjugation to
other elements of the Pauli group.

Note that the ``conjugation-by-$\frac{\pi }{4}A$'' operation $T_{A}\circ
\exp (i\theta B)$ is equivalent to multiplication of $B$ to the left by $iA$
inside the exponent. This is very useful, since the elements of the
normalizer of any stabilizer can always be written as a tensor product of
single qubit Pauli matrices, i.e., as a tensor product of single-body gates.
This is exactly the structure that is suggested by Eq.~(\ref{eq:result2}),
and thus it should allow us to construct $\exp (i\theta \overline{X})$ and $\exp
(i\theta \overline{Z})$ for any Pauli subgroup using at most two-body
interactions. The caveat, however, is that while $\exp (i\theta \overline{X})$
and $\exp (i\theta \overline{Z})$ always preserve the code (since they are in the
normalizer), the operations that generate them from Hamiltonians involving
at most two-body interactions may corrupt the code, as explained in Sec.~\ref
{FTmeaning} above.

Let us then state the challenges ahead: {\em To show how the Hamiltonians} $
\overline{X}$ {\em and} $\overline{Z}$ {\em can be generated using (i) at most
two-body interactions, (ii) fault-tolerantly.}

\subsection{Simple Example: The Subgroup $Q_{4}$}

Let us pause by introducing a simple example illustrating the notion of
universal computation using normalizer elements which are two-body
Hamiltonians. Our example uses a group whose natural structure is such that
the two-body restriction is automatically satisfied. To this end consider
the subgroup $Q_{4}=\{I^{\otimes 4},X^{\otimes 4},Y^{\otimes 4},Z^{\otimes
4}\}$, which we studied in detail in paper 1. It is generated by $K-l=4-l=2$
elements ($X^{\otimes 4},Z^{\otimes 4}$), and therefore encodes $l=2$
qubits, with states given by 
\begin{eqnarray}
|00\rangle_{L} &=&\frac{1}{\sqrt{2}}\left( |0000\rangle +|1111\rangle
\right) \qquad |01\rangle_{L} =\frac{1}{\sqrt{2}}\left( |1001\rangle
+|0110\rangle \right)  \nonumber \\
|10\rangle_{L} &=&\frac{1}{\sqrt{2}}\left( |1100\rangle +|0011\rangle
\right) \qquad |11\rangle_{L} =\frac{1}{\sqrt{2}}\left( |0101\rangle
+|1010\rangle \right) .
\end{eqnarray}
These states are easily seen to be $+1$ eigenstates of $Q_{4}$. The
normalizer in this case contains two $\overline{X}_{i}$ and $\overline{Z}_{i}$
operations, one for each encoded qubit: 
\begin{eqnarray}
\overline{X}_{1} &=&IXXI\qquad \overline{Z}_{1}=ZZII  \nonumber \\
\overline{X}_{2} &=&XXII\qquad \overline{Z}_{2}=IZZI.
\end{eqnarray}
Indeed, we have, for example $\overline{X}_{1}|a,b\rangle _{L}=|1-a,b\rangle _{L}$
and $\overline{Z}_{1}|a,b\rangle _{L}=(-1)^{a}|a,b\rangle _{L}$, so $\overline{X}_{1}$
and $\overline{Z}_{1}$ act, respectively, as a bit flip and a phase flip on the
first encoded qubit. As easily checked, $\overline{X}_{i}$ and $\overline{Z}_{i}$
commute with $Q_{4}$, so that they keep states within the DFS, as should be
the case for normalizer\ elements. As {\em Hamiltonians} $\overline{X}_{i}$ and $
\overline{Z}_{i}$ are valid two-body interactions and hence can be used directly
to generate the encoded $SU(2)$ group on each encoded qubit. That is, $\exp
(i\alpha \overline{X}_{i})$ and $\exp (i\beta \overline{Z}_{i})$ can be combined
directly, with arbitrary values for the angles $\alpha $ and $\beta $, to
produce any operation in $\overline{SU(2)}$ by using the Euler angle
formula. For example, we can construct a rotation about the encoded ${Y}_{i}$
axis by conjugation: $\exp (i\theta \overline{Y}_{i})=\exp (-i\frac{\pi }{4}
\overline{X}_{i})\exp (-i\theta \overline{Z}_{i})\exp (+i\frac{\pi }{4}\overline{X}_{i})$.
We have, therefore, two independent encoded qubits which can be operated
upon seperately by encoded $SU(2)$ operations.

What about coupling between the encoded qubits so that the full $\overline{
SU(4)}$ can be used to do computation? Note that
Hamiltonians like $\overline{Z}_{1}\otimes \overline{Z}_{2}=ZIZI$, which are
two-body on the encoded qubits, can be implemented directly since they are
also two-body on the physical qubits (this is not a generic feature,
however, as discussed in Section~\ref{CSS-many} below). It is a
fundamental theorem of 
universal quantum computation \cite{DiVincenzo:95,Lloyd:95,Barenco:95a} that
the ability to perform $SU(2)$ on two qubits plus the ability to perform 
{\em any} nontrivial two-body {\em Hamiltonian} between these qubits is
universal over the combined $SU(4)$ of these two qubits. Thus we can perform
universal computation on the $Q_{4}$-DFS. In this case the normalizer
elements which perform the $\overline{SU(4)}$ are all two-body Hamiltonians,
and there is no need to apply any new methods in order to perform
fault-tolerant computation which preserves this DFS.

Anticipating the discussion in Section \ref{CNOT}, note that while we have
demonstrated universal computation on a single DFS-block, we have not yet
addressed how to accomplish this when we have clusters of the $Q_{4}$-DFSs.
This, of course, is necessary to scale-up the quantum computer under the $
Q_{4}$-model of decoherence. In order to perform universal fault-tolerant
computation with clusters, we must show that these can be coupled in a
non-trivial manner. Methods for performing non-trivial couplings between
clusters exist for any stabilizer code \cite{Gottesman:97a}. In particular,
the $Q_{4}$-DFS is a Calderbank-Shor-Steane (CSS)-code, whose clusters can
be coupled by performing bit-wise parallel controlled-{\sc not} gates
between two clusters of qubits. This implements as desired an encoded
controlled-{\sc not} between these clusters. In Section \ref{CNOT} we will
discuss what is needed to make this procedure fault-tolerant

$Q_{4}$ is a special case because of the fact that the normalizer elements
are all two-body interactions. In general the normalizer elements will be
many-body interactions and more general techniques are needed, to which we
turn next.

\subsection{Generating $\overline{X}$ and $\overline{Z}$ Using At Most Two-Body
Interactions}

\label{generalXZ}

We now move on to the general case where the normalizer elements are
possibly many-body Pauli operators. Our first task is to show that the
``conjugation-by- $\frac{\pi }{4}A$'' operation $T_{A}\circ \exp (i\theta B)$
can be used to generate any many-body Hamiltonian inside the exponent using
at most two-qubit Hamiltonians. In Section \ref{SU(2)} we show that this is
a fault-tolerant procedure if applied correctly to a DFS.

Suppose the many-body Pauli Hamiltonian $H$ we want to generate is of the
following general form: 
\begin{equation}
H=\sigma _{b}^{\beta }\bigotimes_{j\in {\cal J}}\sigma _{j}^{\alpha _{j}},
\end{equation}
where ${\cal J}$ is some index set and $b\notin {\cal J}$. From Eq.~(\ref
{eq:Hs}) we have at our disposal a single-qubit Hamiltonian $\sigma
_{b}^{\beta }$, and a set of two-qubit Hamiltonians $A_{j}=\sigma
_{b}^{\gamma _{j}}\otimes \sigma _{j}^{\alpha _{j}}$ with $j\in {\cal J}$
and $\gamma _{j}\neq \beta $. We call the $b^{{\rm th}}$ qubit the ``base
qubit''. $A_{j}$ and $\sigma _{b}^{\beta }$ agree on one qubit index but
differ on the Pauli matrix applied to that qubit, so they anticommute: $
\{A_{j},\sigma _{b}^{\beta }\}=0.$ Let ${\cal J}(i)$ denote the $i^{{\rm th}
} $ element in the index set ${\cal J}$. If we use the ``conjugation-by-$
\frac{\pi }{4}A_{{\cal J}(1)}$'' operation about $\exp (i\theta \sigma
_{b}^{\beta })$ [recall Eq.~(\ref{eq:result2})] we obtain:\ 
\begin{equation}
T_{A_{{\cal J}(1)}}\circ \exp (i\theta \sigma _{b}^{\beta })=\exp [i\theta
(i\sigma _{b}^{\gamma _{{\cal J}(1)}}\otimes \sigma _{{\cal J}(1)}^{\alpha _{
{\cal J}(1)}})\sigma _{b}^{\beta }]=\exp [\pm i\theta \sigma _{b}^{\eta
_{1}}\otimes \sigma _{{\cal J}(1)}^{\alpha _{{\cal J}(1)}}],
\end{equation}
where the sign is determined by that of $\varepsilon _{\gamma _{{\cal J}
(1)}\beta \eta _{1}}$, according to the usual rule of multiplication Pauli
matrices: 
\begin{equation}
\sigma ^{\alpha }\sigma ^{\beta }=-i\varepsilon _{\alpha \beta \gamma
}\sigma ^{\gamma }.
\end{equation}
Applying all other ``conjugation-by-$\frac{\pi }{4}A_{{\cal J}(i)}$''
operations, $i=1..|{\cal J}|$, we obtain 
\begin{equation}
T_{A_{{\cal J}(|{\cal J}|)}}\circ \cdots \circ T_{A_{{\cal J}(i)}}\circ \exp
(i\theta \sigma _{b}^{\gamma })=\exp \left( \pm i\theta \bigotimes_{j\in 
{\cal J}}\sigma _{b}^{\eta }\otimes \sigma _{j}^{\alpha _{j}}\right) .
\end{equation}
It is clear that by appropriately choosing the sequence of Pauli matrices,
i.e., the $\alpha _{{\cal J}(i)}$, we can obtain $\eta =\beta $. Further,
conjugating by $-\frac{\pi }{4}$ (instead of $+\frac{\pi }{4}$) allows us to
always adjust the sign in the exponent to $+$. Thus the action of this gate
sequence is to generate the Hamiltonian $H$, as desired: 
\begin{equation}
T_{A_{{\cal J}(|{\cal J}|)}}\circ \cdots \circ T_{A_{{\cal J}(i)}}\circ \exp
(i\theta \sigma _{b}^{\gamma })=\exp \left( i\theta H\right) .
\label{eq:Pauli-prod}
\end{equation}
An example of this type of gate network (analyzed in detail in Section \ref
{Q2X}) is shown in Fig.~\ref{fig:gates-Q2X}. Since the elements of the
normalizer of any stabilizer can always be written as a tensor product of
single qubit Pauli matrices, Eq.~(\ref{eq:Pauli-prod}) gives a constructive
way of generating these normalizer elements as {\em Hamiltonians} (i.e.,
appearing as arguments in the exponent). We have thus met the first
challenge mentioned above: we have shown how to generate the Hamiltonians $
\overline{X}$ and $\overline{Z}$ using at most 2-body interactions. More generally,
Eq.~(\ref{eq:Pauli-prod}) can be considered as a constructive procedure for
generating desired many-body Hamiltonians from given two-body interactions.

Finally, we note that it is perfectly possible to replace the central
single-qubit Hamiltonian with a two-qubit one, specifically by $A_{{\cal J}
(1)}\sigma _{b}^{\gamma }$. This may be more convenient for practical
applications, where control of two-body interactions may be more easily
achievable (as in the case of exchange interactions in quantum dots \cite
{Loss:98}). This change would not affect our fault-tolerance analysis in the
next sections.

\section{Generating Encoded $SU(2)$ Fault-Tolerantly for any Abelian Pauli
Subgroup}

\label{SU(2)}

We are now ready to show how to generate encoded $SU(2)$ operations
fault-tolerantly for any Pauli error-subgroup. Let $Q$ be such a subgroup,
generated by the elements $\{q_{i}\}_{i=1}^{n}$, $|Q|=2^{n}$. Recall that
here these elements play the dual role of errors and of defining the DFS by
fixing its elements. A new (transformed) stabilizer is obtained after each
application of a gate $\exp (i\varphi _{j}A_{j})$. To this sequence of
stabilizers corresponds a sequence of stabilizer-QECCs ${\cal C}_{j}$. Our
strategy will be to find conditions on the Hamiltonians $\{A_{j}\}$ such
that after each gate application, the then current QECC is able to correct
the original $Q$-errors.

Let $Q_{j}$ [$N(Q_{j})$] denote the stabilizer [normalizer] obtained after
application of the gate $U_{j}=\exp (i\varphi _{j}A_{j})$. If $\varphi
_{j}$ is an integer multiple of $\pi/4$ (as we will always assume)
then there are only three mutually exclusive possibilities for the
errors $e\in Q$ (we use the 
notations $e$ and $q$ for members of $Q$ to emphasize the error and
stabilizer element aspects, respectively):

\begin{enumerate}
\item  $e\in Q_{j}$: The error is part of the transformed stabilizer. In
this case the transformed code is immune to $e$ (i.e., the transformed code
is a DFS with respect to $e$), and there is no problem.

\item  $e$ anticommutes with some element of $Q_{j}$: The error is
detectable by the transformed code.

\item  $e\in N(Q_{j})/Q_{j}$ (i.e., $e$ commutes with $Q_{j}$ but is not in
it): The error infiltrated the transformed normalizer. This is a problem
since the error is {\em un}detectable by the transformed code, and acts on
it in a non-trivial manner.
\end{enumerate}

Suppose the errors $e\in Q$ are exclusively of type 1. or 2. Then those that
are of type 2. are not only detectable but also correctable. This is so
because they form a group ($Q$), and therefore any product of two errors is again
either of type 1. or 2., which is exactly the error correction
criterion.\footnote{Note that this is not true for errors in the usual
stabilizer-QECC case, where the errors do not close as a group under
multiplication.} Thus the problematic case is 3., and this is the
case we focus on in order to 
make a prudent choice of Hamiltonians $A_{j}$. To simplify the notation,
from now on we shall denote $N(Q_{j})/Q_{j}$ simply by $N_{j}$ (and by $N$
when $Q_{j}=Q$), and refer to this as the normalizer (without risk of
confusion).

Is there a simple criterion to check whether $e\in N_{j}$? The answer
is contained in the following theorem:

{\it Theorem 2.---} Given are a Pauli-subgroup of errors $Q$, its
normalizer $N$, and 
a sequence of their images $\{Q_{j}\}$ and $\{N_{j}\}$ under conjugation by
unitaries $\{U_{j}\}$. Corresponding to $Q$ is a DFS (code) ${\cal C}$. A
sufficient condition so that no $e\in Q$ is ever in $N_{j}/Q_{j}$ is that
either (i) each $n_{j}\in N_{j}$ equals its source in $N$, or (ii) for each $
n_{j}\in N_{j}$ there exists $m\in N$ such that $\{n_{j},m\}=0$. Then the
transformed codes ${\cal C}_{j}=$ $U_{j}{\cal C}_{j-1}$ (${\cal C}_{1}=U_{1}
{\cal C}$) can always correct the original $Q$-errors.

{\it Proof.---} The normalizer
is, by definition, the set of operations that commute with the stabilizer.
Let us denote this by\ $N=Q^{\prime }$. What is $N^{\prime }$ (the set of
operations that commute with the normalizer)? We have $N^{\prime
}=(Q^{\prime })^{\prime }$, and claim that $(Q^{\prime })^{\prime }=Q$.\footnote{
This is due to a fundamental theorem of $C^{\ast }$-algebras, which states
that if ${\cal A}$ is a $\dagger $-closed algebra (i.e, $a\in {\cal A}$ $
\Longrightarrow $ $a^{\dagger }\in {\cal A}$) which includes the identity,
then it is the commutant of its commutant: ${\cal A}^{\prime \prime }={\cal 
A}$ \cite{Landsman:98a}. These conditions apply to the stabilizer.} In other
words, the only operations that commute with the normalizer are those in the
stabilizer. Now let $n_{j}$ be the image of $n\in N$ after the $j^{{\rm th}}$
transformation. The observation $N^{\prime }=Q$ allows us to exclude case 3.
by checking if, for every $n_{j}\in N_{j}$ (where $n_{j}\neq n$), there
exists $m\in N$ that $n_{j}$ anti-commutes with. To see this, note first
that if $n_{j}=n$ then by definition $n_{j}$ cannot be in $Q$. Secondly, for
an $n_{j}\in N_{j}$ that differs from its source in $N$, assume that it
anti-commutes with some $m\in N$. This implies that $n_{j}$ is not in the
commutant of $N$, and is therefore not in $Q$. If this is true for all $
n_{j}\in N_{j}$ then we have covered the entire new normalizer $N_{j}$ and
not found an element of $Q$ in it. This guarantees that no element of the
original stabilizer $Q$ becomes a member of the new normalizer
$N_{j}$. QED.

Note that if the conditions of the theorem are satisfied then {\em
all} elements of the original stabilizer are excluded from the
transformed normalizer. Therefore also all products of stabilizer
elements are excluded (since the stabilizer is a group), so that all
stabilizer-errors are both detectable and {\em correctable}. 

Below we make repeated use of the result of Theorem 2. The first
application is to show 
how to construct two-body Hamiltonians $\{A_{j}\}$ which can be applied in
succession to produce arbitrary normalizer elements, such that at every
point the theorem is satisfied. To this end we need a basic result from the
theory of stabilizer codes, regarding a standard form for the normalizer. We
then illustrate the general construction with the relatively simple case of
CSS codes, and finally move on to general stabilizer errors.

\subsection{Standard Form of the Normalizer for Stabilizer Codes}

It is shown in \cite{Gottesman:97b} that, due to the fact that the
normalizer is invariant under multiplication by stabilizer elements, the
normalizer of every stabilizer code can be brought into the following
standard form: 
\begin{eqnarray}
\overline{Z}_{j} &=&\mathrel{\mathop{\underbrace{\left( I\otimes \cdots \otimes
I\otimes Z_{j}\otimes I\otimes \cdots \otimes I\right) }}\limits_{l}}\otimes 
\mathrel{\mathop{\underbrace{\left( M_{Z}^j\right) }}\limits_{r}}\otimes 
\mathrel{\mathop{\underbrace{\left( I\otimes \cdots \otimes I\right)
}}\limits_{K-l-r}}  \label{eq:Zbarj} \\
\overline{X}_{j} &=&\mathrel{\mathop{ \underbrace{\left( I\otimes \cdots \otimes
I\otimes X_{j}\otimes I\otimes \cdots \otimes I\right) }}\limits_{l}}\otimes 
\mathrel{\mathop{\underbrace{\left( N_{Z}^j\right) }}\limits_{r}}\otimes 
\mathrel{\mathop{\underbrace{\left( M_{X}^j\right) }}\limits_{K-l-r}}.
\label{eq:Xbarj}
\end{eqnarray}
Here $M_{Z}^j=\otimes _{n\in {\cal Z}_j}Z_{n}$, $N_{Z}^j=\otimes
_{n^{\prime }\in  
{\cal Z}_j^{\prime }}Z_{n^{\prime }}$ and $M_{X}^j=\otimes _{i\in
{\cal X}_j}X_{i}$, where ${\cal Z}_j$, ${\cal Z}^{\prime }_j$ and
${\cal X}_j$ are (possibly empty) index sets of lengths $r$, $r$ and
$K-l-r$ respectively (i.e., $M_{Z}^j$, $N_{Z}^j$ and $M_{X}^j$ are
tensor products of $I$'s and single qubit Pauli $Z$ 
and $X$ matrices, respectively). Recall that $K$ is the number of physical
qubits; $l$ is the number of encoded qubits. The exact form of $M_{Z}^j$, $
N_{Z}^j$ and $M_{X}^j$, as well as the value of the integer $r$, can be found
from the stabilizer \cite{Gottesman:97b}, but is unimportant for our
purposes. We only need the result that for every pair of encoded $Z$ and $X$
operations, acting on the $j^{{\rm th}}$ encoded qubit, it is possible to
express the operations in the blockwise product shown in
Eqs.~(\ref{eq:Zbarj}),(\ref{eq:Xbarj}).

\subsection{CSS-Stabilizer Errors on One Encoded Qubit}

\label{CSS1qubit}

For simplicity, let us now restrict attention to the case of a single
encoded qubit in CSS\ codes, i.e., those codes where every $\overline{Z}$ and $
\overline{X}$ can be written as a product of only $Z$'s and only $X$'s,
respectively. Then, from Eqs.~(\ref{eq:Zbarj}),(\ref{eq:Xbarj}) the standard
form is (dropping the $j$ index):

\begin{eqnarray}
\overline{Z} &=&Z_{1}\otimes M_{Z}\otimes I^{\otimes K-l-r}  \label{eq:ZbarCSS} \\
\overline{X} &=&X_{1}\otimes I^{\otimes r}\otimes M_{X},  \label{eq:XbarCSS}
\end{eqnarray}
i.e., $N_{Z}=I^{\otimes r}$. Our goal is to construct such $\overline{Z}$ and $
\overline{X}$ from single- and two-body Hamiltonians. We shall do this by
starting from the single body Hamiltonians $Z_{1}$ and $X_{1}$, and
conjugating by certain two-body Hamiltonians. The idea is to successively
construct the $Z$'s in $M_{Z}$ and the $X$'s in $M_{X}$. We claim that the
required two-body Hamiltonians have the natural form 
\begin{eqnarray}
A_{n} &=&X_{1}Z_{z_{n}}\qquad z_{n}\in {\cal Z}  \label{eq:An} \\
B_{i} &=&Z_{1}X_{x_{i}}\qquad x_{i}\in {\cal X},  \label{eq:Bi}
\end{eqnarray}
where $n=1..|{\cal Z}|$ and $i=1..|{\cal X}|$, i.e., $A_{n}$ ($B_{i}$) has a 
$Z$ ($X$) in the $n^{{\rm th}}$ ($i^{{\rm th}}$) position of the index set $
{\cal Z}$ (${\cal X}$). If there is an even number of $Z$'s in $\overline{Z}$
then the last Hamiltonian should be taken as $A_{|{\cal Z}|}=X_{1}$ [since
as we show below in that case in the penultimate step we have $Y_{1}\otimes
M_{Z}\otimes I^{\otimes K-l-r}$ for $\overline{Z}$], and similarly for the last $
B_{i}$.\footnote{
For example, suppose $\overline{Z}=Z(ZZZ)(II)$ and $\overline{X}=X(III)(XX)$; then $
A_{1}=X_{1}Z_{2}$, $A_{2}=X_{1}Z_{3}$, $A_{3}=X_{1}Z_{4}$, $A_{4}=X_{1}$ and 
$B_{1}=Z_{1}X_{5}$, $B_{2}=Z_{1}X_{6}$. Then we have: $Z(III)(II)\stackrel{
A_{1}}{\longmapsto }Y(ZII)(II)\stackrel{A_{2}}{\longmapsto }Z(ZZI)(II)
\stackrel{A_{3}}{\longmapsto }Y(ZZZ)(II)\stackrel{A_{4}}{\longmapsto }\overline{Z}
$, and $X(III)(II)\stackrel{B_{1}}{\longmapsto }Y(III)(XI)\stackrel{B_{2}}{
\longmapsto }\overline{X}$.} Note that $[A_{n},\overline{X}]=[B_{i},\overline{Z}]=0$, so that
transforming $\overline{Z}$ does not affect $\overline{X}$, and {\it vice versa}. There
are now two ways to construct $\overline{Z}$ and $\overline{X}$ fault-tolerantly: in
parallel or in series. The parallel implementation has the advantage
that it requires only three basic
steps and thus is very efficient. Its disadvantage is that it may be
hard to implement in practice because it requires simultaneous control
over many qubits. 

\subsubsection{Series Construction}

We assume throughout this discussion that we wish to generate $\exp (i\theta 
\overline{Z})$. The symmetry between $\overline{Z}$ and $\overline{X}$ in the CSS case
implies that our arguments hold for $\exp (i\theta \overline{X})$ as well, with
obvious modifications.

The series construction consists of applying first the sequence of gates $
\{\exp (i\frac{\pi }{4}A_{n})\}_{n=1}^{|{\cal Z}|}$, then the gate $\exp
(i\theta Z_{1})$, and then the reverse sequence of gates $\{\exp (-i\frac{
\pi }{4}A_{n})\}_{n=|{\cal Z}|}^{1}$. An example is shown in Fig.~\ref
{fig:gates-Q2X}. First, as an application of the general Eq.~(\ref
{eq:Pauli-prod}), let us prove that this procedure indeed generates $\exp
(i\theta \overline{Z})$: 
\begin{eqnarray}
T_{A_{1}}\circ T_{A_{2}}\circ \cdots T_{A_{|{\cal Z}|}}\circ \exp (i\theta
Z_{1}) &=&\left[ \bigotimes_{n=1}^{|{\cal Z}|}\exp \left( -i\frac{\pi }{4}
A_{n}\right) \right] \exp (i\theta Z_{1})\left[ \bigotimes_{n=|{\cal Z}
|}^{1}\exp \left( +i\frac{\pi }{4}A_{n}\right) \right]  \nonumber \\
&=&\exp \left[ i\theta (i^{|{\cal Z}|}\prod_{n=1}^{|{\cal Z}|}A_{n}Z_{1})
\right]  \nonumber \\
&=&\exp \left[ (-)^{|{\cal Z}|}i\theta \overline{Z}\right] ,
\end{eqnarray}
where in the first line we used the definition of the ``conjugation-by-$
\frac{\pi }{4}A_{n}$'' operation, in the second the result that this
operation corresponds to multiplication inside the exponent, and in the
third the form in Eq.~(\ref{eq:ZbarCSS}) for $\overline{Z}$. Note that the reason
we have a series of conjugation-by-$\frac{\pi }{4}A_{n}$ operations (as
opposed to trivial identity operations) is that $\{
\prod_{n=1}^{k-1}A_{n}Z_{1},A_{k}\}=0$ $\forall k\leq |{\cal Z}|$. Finally,
we can eliminate the minus sign [if $|{\cal Z}|$ is odd] by changing one of
the $\frac{\pi }{4}$'s to $-\frac{\pi }{4}$.

Next we must demonstrate that the conditions of Theorem 2 are satisfied at
each point in the corresponding circuit in order to guarantee the
fault-tolerance of this implementation. Let us divide the proof into three
parts, by following the transformations of the normalizer elements before
and after the central $\exp (i\theta Z_{1})$ gate, and showing that either $
\overline{Z}$ or $\overline{X}$ anticommutes with the transformed normalizer at each
step along the way.

\underline{Errors before the central gate}:

After application of the first $k$ gates $\{\exp (i\frac{\pi }{4}
A_{n})\}_{n=1}^{k}$, $\overline{Z}$ is transformed to $\overline{Z}^{(k)}\equiv
\prod_{n=1}^{k}A_{n}\overline{Z}$ (we neglect the unimportant factors of $i$ from
now on). From Eq.~(\ref{eq:An}) the product is: 
\begin{equation}
\prod_{n=1}^{k}A_{n}=\left\{ 
\begin{array}{c}
\prod_{n=1}^{k}Z_{z_{n}}\quad {\rm if}\quad k=2l \\ 
X_{1}\prod_{n=1}^{k}Z_{z_{n}}\quad {\rm if}\quad k=2l+1
\end{array}
\right. .
\end{equation}
Therefore, using the standard form: 
\begin{equation}
\left( \prod_{n=1}^{2l}A_{n}\overline{Z}\right) \overline{X}=\prod_{n=1}^{2l}Z_{z_{n}}
\overline{Z}\overline{X}=-\overline{X}\prod_{n=1}^{2l}Z_{z_{n}}\overline{Z}=-\overline{X}\left(
\prod_{n=1}^{2l}A_{n}\overline{Z}\right) ,
\end{equation}
so that $\{\overline{Z}^{(2l)},\overline{X}\}=0$. On the other hand 
\begin{equation}
\left( \prod_{n=1}^{2l+1}A_{n}\overline{Z}\right) \overline{Z}=X_{1}
\prod_{n=1}^{2l+1}Z_{z_{n}}\overline{Z}\overline{Z}=-\overline{Z}X_{1}
\prod_{n=1}^{2l+1}Z_{z_{n}}\overline{Z}=-\overline{Z}\left( \prod_{n=1}^{2l+1}A_{n}
\overline{Z}\right) ,
\end{equation}
so that $\{\overline{Z}^{(2l+1)},\overline{Z}\}=0$. Thus Theorem 2 is satisfied after
each gate application, with $\overline{Z}$ and $\overline{X}$ alternating in the role
of the anticommuting original-normalizer element.

\underline{Error immediately after the central gate}:

At the end of step (i)\ $\overline{Z}$ has been transformed to $Z_{1}$. Since the
central gate ($\theta $-rotation) uses only $Z_{1}$, the transformed $\overline{Z}
$ does not change. Therefore is still anticommutes with the original $\overline{X}
$ and satisfies the criterion of Theorem 2. For the same reason, however, $
\overline{X}$ is transformed by the central gate: 
\begin{equation}
\overline{X}\longmapsto \overline{X}_{\theta }=\overline{X}\cos (2\theta )+i\overline{X}
Z_{1}\sin (2\theta ).
\end{equation}
Thus it suffices to show that $\overline{X}_{\theta }$ anticommutes with
$\overline{Z}$, which is true since $[\overline{Z},Z_{1}]=0$:

\begin{equation}
\overline{X}_{\theta }\overline{Z}=\overline{X}\overline{Z}\cos (2\theta )+i\overline{X}Z_{1}\overline{Z}
\sin (2\theta )=-\overline{Z}\overline{X}\cos (2\theta )-i\overline{Z}\overline{X}Z_{1}\sin
(2\theta )=-\overline{Z}\overline{X}_{\theta }.
\end{equation}

\underline{Errors after the central gate}:

After application of the first $k^{\prime }$ inverse gates $\{\exp (-i\frac{
\pi }{4}A_{n})\}_{n=|{\cal Z}|}^{|{\cal Z}|-k^{\prime }+1}$, $Z_{1}$ is
transformed to $\overline{Z}^{\prime (k)}\equiv \prod_{n=1}^{k^{\prime }}A_{n}
\overline{Z}$. Therefore the same reasoning as in (i)\ applies to $\overline{Z}
^{\prime (k)}$. As for $\overline{X}$ (which is now $\overline{X}_{\theta }$), the $
\overline{X}\cos (2\theta )$ component commutes with the inverse gates $\exp (-i
\frac{\pi }{4}A_{n})$ so that it does not change. The $i\overline{X}Z_{1}\sin
(2\theta )$ component, however, anticommutes with the inverse gates $\exp (-i
\frac{\pi }{4}A_{n})$. Therefore it flips back and forth between $i\overline{X}
Z_{1}\sin (2\theta )$ and $i\overline{X}Y_{1}\sin (2\theta )$. These terms
anticommute with the original $\overline{Z}$ and $\overline{Y}$, respectively. But so
does the $\overline{X}\cos (2\theta )$ component, so their sum anticommutes
alternately with the original $\overline{Z}$ and $\overline{Y}$.

We conclude that Theorem 2 is satisfied at each stage of the circuit.
Therefore the series construction is indeed fault-tolerant. Of course, this
fault-tolerance is achieved in practice by supplementing the circuit with
error-detection and correction procedures after each gate (the parallel
construction discussed next is much more economical for this reason). We
discuss this issue in Section \ref{Measurements}.

\subsubsection{Parallel Construction}

\label{parallel}

Since the $A_{n}$ ($B_{i}$) all commute, the corresponding gates can also be
implemented {\em in parallel}. That is, 
\begin{eqnarray}
U_{A} &\equiv &\bigotimes_{n\in {\cal Z}}\exp \left( i\frac{\pi }{4}
A_{n}\right) =\exp \left( i\frac{\pi }{4}\sum_{n\in {\cal Z}}A_{n}\right) 
\nonumber \\
U_{B} &\equiv &\bigotimes_{i\in {\cal X}}\exp \left( i\frac{\pi }{4}
B_{i}\right) =\exp \left( i\frac{\pi }{4}\sum_{i\in {\cal X}}B_{i}\right)
\end{eqnarray}
can be used as parallel gates in our circuit (see Fig.~\ref{fig:parallel}
for an example). To see directly that this circuit really does implement the
normalizer gate $\exp (i\theta \overline{Z})$ [or $\exp (i\theta \overline{X})$],
observe that, by definition $\{A_{n},Z_{1}\}=\{B_{i},X_{1}\}=0$ for all $n$
and $i$. This means that conjugation of $Z_{1}$ by $U_{A}$ will act as
multiplication by $\prod_{n\in {\cal Z}}A_{n}$ and thus transform $Z_{1}$ to 
$\overline{Z}$ (without changing $X_{1}$). The same is true for $X_{1}$ by
changing $Z$'s to $X$'s and $U_{A}$ to $U_{B}$. Therefore $
U_{A}Z_{1}U_{A}^{\dagger }=\overline{Z}$ and $U_{B}X_{1}U_{B}^{\dagger }=\overline{X}$
, from which follows immediately by Taylor expansion that: 
\begin{eqnarray}
U_{A}\exp (i\theta Z_{1})U_{A}^{\dagger } &=&\exp (i\theta \overline{Z}) 
\nonumber \\
U_{B}\exp (i\theta X_{1})U_{B}^{\dagger } &=&\exp (i\theta \overline{X}).
\end{eqnarray}
This too is a fault-tolerant construction. The reason is that it corresponds
to looking at the series construction just at the following three points:
right before the central gate, right after the central gate, and the end.

\subsection{Example: The Subgroup $Q_{2X}$}

\label{Q2X}

As an example with a many-body normalizer element, consider the Pauli
subgroup/stabilizer generated by the errors $XXII$, $IXXI$, $IIXX$: 
\begin{equation}
Q_{2X}=\{IIII,XXII,XIIX,IIXX,XIXI,IXXI,IXIX,XXXX\}.  \label{eq:Q2x}
\end{equation}
It describes a physically interesting error-model, of bit-flip errors which
act on all pairs of nearest neighbor qubits. This situation is of interest,
e.g., when decoherence results from spin-rotation coupling in a dipolar
Hamiltonian, typical in NMR \cite{Slichter:96}: 
\begin{equation}
H_{I}=\sum_{j,k}\frac{\gamma _{j}\gamma _{k}}{r_{jk}^{3}}\left[ {\bf \sigma }
_{j}\cdot {\bf \sigma }_{k}-3\left( {\bf \sigma }_{j}\cdot {\bf r}
_{jk}\right) \left( {\bf \sigma }_{k}\cdot {\bf r}_{jk}\right) \right] .
\end{equation}
Here $\gamma _{j}$ is the gyromagnetic ratio of spin $j$, $r_{jk}$ is the
distance beween spins $j$ and $k$. In the anistropic case (e.g., a liquid
crystal) this can be rewritten as 
\begin{equation}
H_{I}=\sum_{j,k}\frac{\gamma _{j}\gamma _{k}}{r_{jk}^{3}}\sum_{\alpha ,\beta
=-1}^{1}g_{jk}^{\alpha \beta }\left( \sigma _{j}^{\alpha }\otimes \sigma
_{k}^{\beta }\right) Y_{2}^{-\alpha -\beta },
\end{equation}
where $Y_{l}^{m}$ are the spherical harmonics and $g_{jk}^{\alpha \beta }$
is the anistropy tensor. When $g_{jk}^{\alpha \beta }=\delta _{\alpha
0}\delta _{\beta 0}g_{jk}$ only the $\sigma _{j}^{z}\otimes \sigma _{k}^{z}$
terms remain (coupled to $Y_{2}^{0}$), which leads to decoherence described
by the subgroup $Q_{2Z}$ (defined similarly to $Q_{2X}$), analyzed in paper
1.

To find the DFS under $Q_{2X}$, we construct in accordance with the
techniques of paper 1 the projector $P=\frac{1}{8}\sum_{i}q_{i}$
(corresponding to the identity irrep of $Q_{2X})$, where the sum is over all 
$q_{i}\in Q_{2X}$. Applying this projector to an arbitrary initial state we
find a 2-dimensional DFS, spanned by the states 
\begin{eqnarray}
|0_{L}\rangle &=&\left( |0000\rangle +|0011\rangle +|0101\rangle
+|0110\rangle +|1001\rangle +|1010\rangle +|1100\rangle +|1111\rangle
\right) /\sqrt{8},  \nonumber \\
|1_{L}\rangle &=&\left( |0001\rangle +|0010\rangle +|0100\rangle
+|0111\rangle +|1000\rangle +|1011\rangle +|1101\rangle +|1110\rangle
\right) /\sqrt{8}.  \label{eq:Q2X-states}
\end{eqnarray}
This DFS thus encodes a full qubit.

Since for $Q_{2X}$ there is just one encoded qubit, we expect to find just
one $\overline{X}$ and one $\overline{Z}$. In the case of $Q_{2X}$ it is easily
verified that the normalizer is generated by 
\begin{eqnarray}
\overline{X} &=&XIII,\quad  \nonumber \\
\overline{Z} &=&ZZZZ.
\end{eqnarray}
$\overline{X}$ is already a single-body Hamiltonian and therefore can be
implemented directly. Let us show how $\overline{Z}$ can be implemented as a
Hamiltonian using at most two-body interactions.

Note that $Q_{2X}$ supports a CSS code. Comparing the above expressions for $
\overline{Z}$ to the standard form for CSS normalizer elements [Eq.~(\ref
{eq:ZbarCSS})], we have $M_{Z}=Z_{2}Z_{3}Z_{4}$ and $M_{X}=\emptyset $.
Therefore, from the recipe of Eq.~(\ref{eq:An}): $A_{n}=X_{1}Z_{n+1}$ for $
n=1..3$, while $A_{4}=X_{1}$. The series-circuit implementing $\exp (i\theta 
\overline{Z})$ thus has the form depicted in Fig.~\ref{fig:gates-Q2X}. The
parallel version of the same circuit is shown in Fig.~\ref{fig:parallel}. To
verify directly that these circuits indeed implement $\exp (i\theta 
\overline{Z})$ use Eq.~(\ref{eq:Pauli-prod}) and choose the base qubit to be
the first qubit. Then: 
\begin{equation}
T_{XZII}\circ T_{XIZI}\circ T_{XIIZ}\circ T_{XIII}\circ \exp (i\theta
ZIII)=\exp (i\theta ZZZZ).
\end{equation}
As required, this is an implementation that uses at most two-body
interactions.

Fig.~\ref{fig:gates-Q2X} also shows the transformed $\overline{Z}$ at each point,
and directly below the original normalizer element ($\overline{X}$ or $\overline{Z}$)
that this transformed normalizer element anticommutes with. This verifies
that the circuit is indeed a fault-tolerant implementation of $\exp (i\theta 
\overline{Z})$ for $Q_{2X}$.

\subsection{CSS-Stabilizer Errors on Multiple Encoded Qubits}
\label{CSS-many}

The CSS case of more than one encoded qubit is a simple extension of the
single encoded qubit case discussed above. From Eqs.~(\ref{eq:Zbarj}),(\ref
{eq:Xbarj}) the standard form for a CSS code is now:

\begin{eqnarray}
\overline{Z}_{j} &=&Z_{j}\otimes M_{Z}^j\otimes I^{\otimes K-l-r}
\label{eq:ZjbarCSS} \\
\overline{X}_{j} &=&X_{j}\otimes I^{\otimes r}\otimes M_{X}^j.
\label{eq:XjbarCSS}
\end{eqnarray}
Operations on different encoded qubits $j$,$j^{\prime }$
commute. Therefore the single encoded qubit 
constructions still holds when the Hamiltonians are modified to read 
\begin{eqnarray}
A_{n}^{(j)} &=&X_{j}Z_{z_{n}}\qquad z_{n}\in {\cal Z}_j   \\
B_{i}^{(j)} &=&Z_{j}X_{x_{i}}\qquad x_{i}\in {\cal X}_j.
\end{eqnarray}
As is easily checked, the entire proof for the single encoded qubit case
carries through when the base qubit becomes physical qubit number $j$
instead of number $1$. This thus allows us to fault-tolerantly implement $
\overline{SU(2)}^{\otimes l}$ on all $l$ encoded qubits.

To couple encoded
qubits within the same block [thus generating $\overline{SU(2^{l})}$],
one could 
use a standard trick from stabilizer theory \cite{Gottesman:97a}, using an
auxiliary block to swap information into and out of. This transversal
operation involves applying encoded controlled-{\sc NOT} operations, which
we treat in Section \ref{CNOT} below. In that Section we also show how
coupling multiple encoded qubits can be achieved directly, without
resorting to an auxiliary block.

\subsection{General Stabilizer Errors}

The entire analysis for the CSS case carries through in the general
stabilizer case for the implementation of $\exp (i\theta \overline{Z})$,
since $\overline{Z}$ remains unchanged [recall Eq.~(\ref{eq:Zbarj})].
However, the encoded $X$ operation now includes the additional block $N_{Z}$: $\overline{X}=X_{1}\otimes N_{Z}\otimes M_{X}$ [Eq.~(\ref{eq:Xbarj})].
Therefore to generate this operation we must include a new set of
Hamiltonians: 
\begin{equation}
C_{n^{\prime }}=Z_{1}Z_{n^{\prime }}\qquad n^{\prime }\in {\cal Z}^{\prime }.
\end{equation}
If there is an even number of $Z$'s in $\overline{Z}$ then the last Hamiltonian
should be taken as $C_{|{\cal Z}^{\prime }|}=Z_{1}$. We now need to repeat
the analysis for the generation of $\exp (i\theta \overline{X})$. Again,
there is a series and a parallel construction. Since the $C_{n^{\prime
}}$ and $B_{i}$ all commute, the gate 
\begin{equation}
U_{BC}\equiv U_{B}\otimes U_{C}=\left[ \bigotimes_{i\in 
{\cal X}}\exp \left( i\frac{\pi }{4}B_{i}\right) \right] \otimes \left[
\bigotimes_{n^{\prime }\in {\cal Z}^{\prime }}\exp \left( i\frac{\pi }{4}
C_{n^{\prime }}\right) \right] =\exp \left[ i\frac{\pi }{4}\left(
\sum_{i\in {\cal X}}B_{i}+\sum_{n^{\prime }\in {\cal Z}^{\prime
}}C_{n^{\prime }}\right) \right]
\end{equation}
can be implemented in parallel. Conjugation of $\exp (i\theta X_{1})$ by $
U_{BC}$ will yield $\exp (i\theta \overline{X})$ by Eq.~(\ref
{eq:Pauli-prod}), since $\{X_{1},B_{i}\}=\{X_{1},C_{n^{\prime
}}\}=0$. It is further straightforward to check that this is a
fault-tolerant implementation, since the arguments used in the case of a
single encoded CSS qubit are still valid here.

We are thus left to check only the series construction. Here the only new
element is that we must make sure that the application of the $C_{n^{\prime
}}$ Hamiltonians does not allow for undetectable errors to take place.
Apart from this everything is the same as in the CSS case. Now, after
application of the first $k$ gates $\{\exp (i\frac{\pi }{4}C_{n^{\prime
}}\}_{n^{\prime }=1}^{k}$, $\overline{X}$ is transformed to
$\overline{X}^{(k)}\equiv \prod_{n^{\prime }=1}^{k}C_{n^{\prime
}}\overline{X}$. 
This product is: 
\begin{equation}
\prod_{n^{\prime }=1}^{k}C_{n^{\prime }}=\left\{ 
\begin{array}{c}
\prod_{n^{\prime }\in {\cal Z}_{k}^{\prime }}Z_{n^{\prime }}\quad {\rm if}
\quad k=2l \\ 
Z_{1}\prod_{n^{\prime }\in {\cal Z}_{k}^{^{\prime }}}Z_{n^{\prime }}\quad 
{\rm if}\quad k=2l+1
\end{array}
\right. ,
\end{equation}
where ${\cal Z}_{k}^{^{\prime }}$ are the first $k$ elements of the index
set ${\cal Z}^{\prime }$. Therefore 
\begin{equation}
\{\overline{X}^{(k)},\overline{Z}\}=\left[ \left( Z_{1}\right)
^{k}\prod_{n^{\prime }\in {\cal Z}_{k}^{^{\prime }}}Z_{n^{\prime }}\overline{X}
\right] \overline{Z}+\overline{Z}\left[ \left( Z_{1}\right)
^{k}\prod_{n^{\prime }\in {\cal Z}_{k}^{^{\prime }}}Z_{n^{\prime }}\overline{X}
\right] =\left[ \left( Z_{1}\right) ^{k}\prod_{n^{\prime }\in {\cal Z}
_{k}^{^{\prime }}}Z_{n^{\prime }}\right] \{\overline{X},\overline{Z}\}=0.
\end{equation}
Thus Theorem 2 is satisfied after each $C_{n^{\prime }}$-gate
application, with $\overline{Z}$ playing the role of the anticommuting
original-normalizer element. This means that use of the Hamiltonians $
C_{n^{\prime }}$ does not spoil the fault tolerance of the circuit. We
know from the calculations in the single encoded qubit case that the rest of
the circuit is also fault tolerant. Hence we can conclude at this point that
our method of constructing normalizer elements is fault tolerant for any
stabilizer code.

\subsection{Summary}

Let us recapitulate the main result of this section. Given a set of errors
corresponding to some Abelian subgroup of the Pauli group (i.e., a
stabilizer), there is a DFS which is immune against these errors. We have
shown how to implement arbitrary encoded $SU(2)$ operations on this class of
DFSs. To do so, we gave an explicit construction of encoded $\sigma _{x}$
and $\sigma _{z}$ operations, which together span encoded $SU(2)$'s for each
DFS-qubit. The construction involves turning on and off a series of one- and
two-body Hamiltonians for a specific durations. Each such operation takes
the encoded states outside of the DFS. However, our construction guarantees
that the errors always remain correctable by the code formed by the
transformed states. That is, these states form a QECC with respect to the
Pauli subgroup errors. Therefore, our construction works by supplementing
the unitary gates executing the encoded $\sigma _{x}$ and $\sigma _{z}$
operations by appropriate error correction procedures. To complete the
construction, we still need to show how to execute encoded two-body gates,
and how to fault-tolerantly measure the error syndrome. This is the subject
of the next two sections.

\section{Encoded Controlled-{\sc NOT}}

\label{CNOT}

The unitary controlled-{\sc NOT} ({\sc CNOT)} operation from the first qubit
(``control qubit)\ to the second qubit (``target qubit'') can be written in
the basis of $\sigma _{z}$ eigenstates as: 
\begin{equation}
U_{\text{{\sc CNOT}}}=\left( 
\begin{array}{cc}
I & 0 \\ 
0 & X
\end{array}
\right)
\end{equation}
(where each entry is a $2\times 2$ matrix). Since we are working in the
Heisenberg picture it is useful to consider how two-qubit operators
transform under {\sc CNOT}. For example, 
\begin{equation}
X\otimes I\longmapsto U_{\text{{\sc CNOT}}}\left( X\otimes I\right) U_{\text{
{\sc CNOT}}}^{\dagger }=\left( 
\begin{array}{cc}
I & 0 \\ 
0 & X
\end{array}
\right) \left( 
\begin{array}{cc}
0 & I \\ 
I & 0
\end{array}
\right) \left( 
\begin{array}{cc}
I & 0 \\ 
0 & X
\end{array}
\right) =\left( 
\begin{array}{cc}
0 & X \\ 
X & 0
\end{array}
\right) .
\end{equation}
As is simple to verify, the full transformation table is: 
\begin{eqnarray}
X\otimes I &\longmapsto &X\otimes X  \nonumber \\
I\otimes X &\longmapsto &I\otimes X  \nonumber \\
Z\otimes I &\longmapsto &Z\otimes I  \nonumber \\
I\otimes Z &\longmapsto &Z\otimes Z.  \label{eq:transf-table}
\end{eqnarray}
Since $U(A\otimes B)U^{\dagger }=U(A\otimes I)U^{\dagger }U(I\otimes
B)U^{\dagger }$, the rest of the transformations under {\sc CNOT} follow
simply by taking appropriate products of the above, e.g., $X\otimes Z=\left(
X\otimes I\right) \left( I\otimes Z\right) \longmapsto \left( X\otimes
X\right) \left( Z\otimes Z\right) =-Y\otimes Y$. We need to show how to
fault-tolerantly construct an encoded {\sc CNOT} operation for the DFS
corresponding to a given Pauli subgroup of errors.

\subsection{CSS-Stabilizer Errors}

It is well known that a bitwise {\sc CNOT}\ gate between physical qubits in
different blocks is an operation that preserves any CSS code, and acts as
the encoded {\sc CNOT}\ gate between the blocks encoding different qubits 
\cite{Gottesman:97a}. However, this is true only at the {\em conclusion} of
the operation, i.e., after all the bitwise operations have been applied.
During the execution of the bitwise operations the codewords are exposed to
errors. To demonstrate this, consider the transformation of the normalizer
elements of a CSS code. Let {\sc CNOT}$_{j_{A},j_{B}}$ denote the {\sc CNOT}
operation from control qubit $j$ (in the first block $A$) to target qubit $j$
(in the second block $B$). For definiteness let us consider the
transformation of $\overline{X}_{j}\otimes I^{\otimes K}$ under bitwise {\sc CNOT}'s. Then, because of the standard form for $\overline{X}_{j}$, the first {\sc CNOT} operation is applied from control qubit $j$, and subsequent {\sc CNOT}'s
from control qubits determined by the index set ${\cal X}$, i.e., acting on
pairs of physical qubits at positions $\{(i_{A},i_{B})\}_{i\in {\cal X}}$.
Using Eqs.~(\ref{eq:ZjbarCSS}),(\ref{eq:XjbarCSS}) for $\overline{Z}_{j}$ and $
\overline{X}_{j}$, and the transformation table of Eq.~(\ref{eq:transf-table}),
we find: 
\begin{eqnarray}
\overline{X}_{j}\otimes I^{\otimes K} &=&[X_{j}\otimes I^{\otimes r}\otimes
M_{X}]\otimes \lbrack I^{\otimes l}\otimes I^{\otimes r}\otimes I^{\otimes
K-l-r}]  \nonumber \\
&\stackrel{\text{{\sc CNOT}}_{j_{A},j_{B}}}{\longmapsto }&[X_{j}\otimes
I^{\otimes r}\otimes M_{X}]\otimes \lbrack X_{j}\otimes I^{\otimes r}\otimes
I^{\otimes K-l-r}]  \nonumber \\
&\stackrel{\text{{\sc CNOT}}_{i_{1A},i_{1B}}}{\longmapsto }&[X_{j}\otimes
I^{\otimes r}\otimes M_{X}]\otimes \lbrack X_{j}\otimes I^{\otimes r}\otimes
X_{i_{1}}]  \nonumber \\
&\longmapsto &...\longmapsto \lbrack X_{j}\otimes I^{\otimes r}\otimes
M_{X}]\otimes \lbrack X_{j}\otimes I^{\otimes r}\otimes M_{X}]=\overline{X}
_{j}\otimes \overline{X}_{j}.  \label{eq:XbarI}
\end{eqnarray}
Similarly one can check that the rest of the transformations of
Eq.(~\ref{eq:transf-table}) are satisfied at the encoded level. Therefore this calculation demonstrates that the full bitwise {\sc CNOT} gate indeed
acts as an {\em encoded }{\sc CNOT} operation, since it transforms encoded
normalizer operations according to the transformation rules of {\sc CNOT},
as per Eq.~(\ref{eq:transf-table}). In our context this implies that given a
certain Pauli subgroup of errors, application of the full bitwise {\sc CNOT}
gate will implement the $\overline{\text{{\sc CNOT}}}$ gate on the DFS in
a way which keeps the codewords inside the DFS at the end of the operation.
However, as in the $\overline{SU(2)}$ case, this is not true at intermediate
steps, meaning that the code leaves the DFS.\footnote{
Note that this is equally true for stabilizer-QECCs, which are thus exposed
to errors during gate execution.} As in the $\overline{SU(2)}$ case, we must
check that the original errors are still correctable at each intermediate
step. Theorem 2 will still apply if error correction procedures are
implemented on each block separately, after each bitwise {\sc CNOT}\
operation (since the blocks are only coupled during the execution of the 
{\sc CNOT}). Therefore we need to check that for each block in which the
normalizer changed, there exists an element in the original normalizer that
anticommutes with the transformed normalizer. It is easy to see from
Eq.~(\ref{eq:XbarI}) that $\overline{X}_{j}$ does not change in the first
block, and 
the sequence of transformed $\overline{X}_{j}$'s in the second block anticommutes
with $\overline{Z}_{j}$ at every step. Therefore error correction is possible at
each intermediate step.

To complete the construction it is necessary to check that the remaining
normalizer elements are appropriately transformed. Repeating the calculation
of Eq.~(\ref{eq:XbarI}) it is straightforward to check that this is true,
namely: 
\begin{eqnarray}
I^{\otimes K}\otimes \overline{X}_{j} &\longmapsto &I^{\otimes K}\otimes \overline{X}
_{j}  \nonumber \\
\overline{Z}_{j}\otimes I^{\otimes K} &\longmapsto &\overline{Z}_{j}\otimes I^{\otimes
K}  \nonumber \\
I^{\otimes K}\otimes \overline{Z}_{j} &\longmapsto &\overline{Z}_{j}\otimes \overline{Z}
_{j},
\end{eqnarray}
with $I^{\otimes K}\otimes \overline{X}_{j}$ and $\overline{Z}_{j}\otimes I^{\otimes
K} $ invariant under the bitwise {\sc CNOT}'s (thus requiring no error
correction), and the transformed $I^{\otimes K}\otimes \overline{Z}_{j}$
anticommuting at each step with the original $\overline{X}_{j}$.

This completes our demonstration that a $\overline{\text{{\sc CNOT}}}$ gate
can be implemented fault-toleranty using bitwise {\sc CNOT}'s in the CSS
case.

\subsection{General Stabilizer Errors}

In the non-CSS case the bitwise {\sc CNOT} does not act as a $\overline{
\text{{\sc CNOT}}}$. One quick way to realize this is to note that since $
X\otimes I\longmapsto X\otimes X$, by unitarity $X\otimes X\longmapsto
X\otimes I$, but this is not the case at the encoded level: 
\begin{eqnarray*}
\overline{X}\otimes \overline{X} &=&[X_{1}\otimes N_{Z}\otimes M_{X}]\otimes \lbrack
X_{K+1}\otimes N_{Z}\otimes M_{X}] \\
&\longmapsto &[X_{1}\otimes I^{\otimes r}\otimes M_{X}]\otimes \lbrack
I_{K+1}\otimes N_{Z}\otimes I^{\otimes K-l-r}]\neq \overline{X}\otimes I^{\otimes
K}.
\end{eqnarray*}
Thus a different implementation of the $\overline{\text{{\sc CNOT}}}$ is
needed. Now, it is clear that if the product of stabilizers for different
blocks (each encoding one qubit or more) is mapped to itself at the end of
the $\overline{\text{{\sc CNOT}}}$ implementation, then the stabilizer
errors will not have changed, the DFS qubits will not have changed, and thus
the DFS-code still offers protection against the stabilizer errors.
Gottesman \cite{Gottesman:97a} has given such an implementation of the $
\overline{\text{{\sc CNOT}}}$ for arbitrary stabilizer codes. It uses
transformations involving $4$ blocks at a time, where two blocks serve as
ancillas and are discarded after a measurement at the end of the
implementation. We will not repeat this analysis here -- the interested
reader is referred to p.133 of \cite{Gottesman:97a} for details. The faster
the gate sequence implementing this $\overline{\text{{\sc CNOT}}}$ is
executed compared to the timescale for the errors to appear, the higher the
probability that the code will not be taken outside of the DFS. However, as
shown in Appendix \ref{app:A}, the gate sequence (Fig. 2 of \cite
{Gottesman:97a}) does not have the property we have been able to demonstrate
above for all our constructions, i.e., it allows for errors to become part
of the transformed normalizer. Therefore we cannot use this construction.
Instead we now introduce a different construction for the $\overline{\text{
{\sc CNOT}}}$, in the spirit of what we have done above for the $\overline{
SU(2)}$ operations.

Consider two blocks $A$ and $B$ encoding one DFS qubit each. We already know
how to implement $\exp (i\theta I_{A}\otimes \overline{X}_{B})$. Suppose one
can also implement $\exp (i\theta \overline{Z}_{A}\otimes \overline{X}_{B})$. Then by use of the Trotter formula $\exp [i(t_{1}O_{1}+t_{2}O_{2})/n]${$
=\lim_{n\rightarrow \infty }\left[ \exp \left( i\frac{t_{1}}{n}O_{1}\right)
\exp \left( i\frac{t_{2}}{n}O_{2}\right) \right] ^{n}$ \cite{Bhatia}, or its
short-time approximation} 
\begin{equation}
\exp [it(O_{1}+O_{2})/n]=\exp [itO_{1}/n]\exp [itO_{2}/n]+O(n^{-2})
\end{equation}
valid for arbitrary operators $O_{1}$ and $O_{2}$, we can form, to any
desired accuracy 
\begin{equation}
\exp [i\theta (I_{A}\otimes \overline{X}_{B}-\overline{Z}_{A}\otimes 
\overline{X}_{B})/2]=\left( 
\begin{array}{cc}
I & 0 \\ 
0 & \exp (i\theta \overline{X}_{B})
\end{array}
\right) .
\end{equation}
For $\theta =\pi /2$ this is the $\overline{\text{{\sc CNOT}}}$ operation
between the two blocks. Thus our problem reduces to showing how $\exp
(i\theta \overline{Z}_{A}\otimes \overline{X}_{B})$ can be implemented
fault-tolerantly for arbitrary stabilizer DFSs.

Consider the circuit shown in Fig~\ref{fig:CNOT}. It describes the
implementation of $\overline{Z}$ and $\overline{X}$ operations, as in the $
\overline{SU(2)}$ case, with the difference that the single-body central
gates have been replaced with a two-body gate, generated by the Hamiltonian $
H_{AB}=Z_{1}^{A}\otimes X_{1}^{B}$ (here $A$ and $B$ are the two blocks and
the subscript $1$ indicates the first physical qubit in each block). By the $
\overline{SU(2)}$ construction we have that $U_{A}Z_{1}^{A}U_{A}^{\dagger }=
\overline{Z}_{A}$ and $U_{B}X_{1}^{B}U_{B}^{\dagger }=\overline{X}_{B}$
(recall Sec.~\ref{parallel}). Therefore, using the fact that for any
non-singular matrix $M$ the equality $M\exp (H)M^{-1}=\exp (MHM^{-1})$
holds, the gates in Fig.~\ref{fig:CNOT} yield: 
\begin{eqnarray}
\left( U_{A}\otimes U_{B}\right) \exp (i\theta H_{AB})\left( U_{A}^{\dagger
}\otimes U_{B}^{\dagger }\right) &=&\exp \left[ i\theta \left( U_{A}\otimes
U_{B}\right) H_{AB}\left( U_{A}^{\dagger }\otimes U_{B}^{\dagger }\right) 
\right]  \nonumber \\
&=&\exp \left[ i\theta \left( U_{A}Z_{1}^{A}U_{A}^{\dagger }\right) \otimes
\left( U_{B}X_{1}^{B}U_{B}^{\dagger }\right) \right]  \nonumber \\
&=&\exp \left( i\theta \overline{Z}_{A}\otimes \overline{X}_{B}\right) ,
\end{eqnarray}
as desired.

It remains to verify that this is a fault-tolerant construction. The only
difference compared to the $\overline{SU(2)}$ construction above is the fact
that we are now using a {\em two}-body central Hamiltonian. It is reasonable
to assume that if the system can couple the two blocks connected by this
Hamiltonian, then so can the environment. Therefore instead of considering
the error subgroups $Q_{A}$ and $Q_{B}$ separately, we must now consider the
new error subgroup $Q_{A}\times Q_{B}$. But then the appropriate normalizer
is $N_{AB}=N_{A}\times N_{B}$, and the sequence of transformed normalizers
satisfy $N_{AB,j}=N_{A,j}\times N_{B,j}$. This makes the fault-tolerance
verification task very simple:\ We already checked in our $\overline{SU(2)}$
discussion that Theorem 2 is satisfied for each block separately. Now,
clearly both $N_{A}\otimes I_{B},I_{A}\otimes N_{B}\in N_{AB}$. Therefore,
since for every transformed normalizer element in $N_{A,j}$ [$N_{B,j}$]
there is an anticommuting element in the original normalizer $N_{A}$
[$N_{B}$], it follows that $N_{A}\otimes I_{B}$ [$I_{A}\otimes N_{B}$] will
correspondingly anticommute with the elements of $N_{AB,j}$. This means that
Theorem 2 is satisfied also for the combination of blocks $A$ and $B$, and
fault-tolerance is guaranteed as in the $\overline{SU(2)}$ case.

As promised in Section~\ref{CSS-many}, the construction presented here
also applies to multiple qubits encoded into a single block. To see
this, consider the case of two encoded qubits in the same block, and
let us show that we can generate $\exp(i\theta \overline{Z}_1 \otimes
\overline{Z}_2)$ between them. This coupling, together with single
encoded-qubit operations, suffices to generate $\overline{SU(2^{l})}$
(for $l$ encoded qubits in a block). Now, from the standard form we
have:
\begin{eqnarray}
\overline{Z}_{1} &=&Z_{1}\otimes M_{Z}^1\otimes I^{\otimes K-l-r} \\
\overline{Z}_{2} &=&Z_{2}\otimes M_{Z}^2\otimes I^{\otimes K-l-r}.
\end{eqnarray}
Let $\overline{Z}_{1} = U_{1}Z_1 U_{1}^{\dagger}$ and $\overline{Z}_{2}
= U_{2}Z_1 U_{2}^{\dagger}$. Note that $[U_{1},Z_2]=[U_{2},Z_1]=0$
since $U_{1(2)}$ contains $X_{2(1)}$. For the same reason also
$[U_1,U_2]=0$. Therefore:
\begin{eqnarray}
\left( U_{1}\otimes U_{2}\right) \exp (i\theta Z_{1}\otimes Z_2)
\left( U_{1}^{\dagger }\otimes U_{2}^{\dagger }\right) &=&\exp \left[
i\theta \left( U_{1}\otimes U_{2}\right) Z_{1}\otimes Z_2\left(
U_{1}^{\dagger }\otimes U_{2}^{\dagger }\right) \right]  \nonumber \\
&=&\exp \left[ i\theta \left( U_{1}Z_{1}U_{1}^{\dagger }\right) \otimes
\left( U_{2}Z_{2}U_{2}^{\dagger }\right) \right]  \nonumber \\
&=&\exp \left( i\theta \overline{Z}_{1}\otimes \overline{Z}_{2}\right) .
\end{eqnarray}

The same idea can be used to implement $\overline{\text{{\sc CNOT}}}$
between multiple qubits encoded into a single block. We have thus provided a
fault-tolerant implementation of $\overline{\text{{\sc CNOT}}}$ for any
stabilizer DFS.

\section{Fault Tolerant Measurement of the Error Syndrome}

\label{Measurements}

So far we have taken for granted that error detection and correction is
possible in between gate applications. We now complete our discussion by
showing that it is indeed possible to do so fault-tolerantly. This requires
the ability to measure the sequence of transformed stabilizer generators in
a manner that does not introduce new errors in a catastrophic way. To
accomplish this fault-tolerant measurement we follow, with some
modifications, the usual stabilizer construction \cite{Gottesman:97b}.

Let us recall the basics of measurement within stabilizer theory. A DFS
state $|\psi \rangle $ in the stabilizer $Q$ is a $+1$ eigenstate of all
elements of $Q$. An error $e$ is an operator that anticommutes with at least
one element of the stabilizer $Q$, say $q$. If $|\psi \rangle \in Q$ then $
qe|\psi \rangle =-eq|\psi \rangle =-e|\psi \rangle $, so that $e|\psi
\rangle $ is an eigenstate of $q$ with eigenvalue $-1$. Therefore each
generator measurement that returns the eigenvalue $+1$ indicates that no
error has occured, while each $-1$ result indicates an error, which can be
fixed by applying the error $e$ to the state. The sequence of $\pm 1$'s that
results from measuring all stabilizer generators is called the ``error
syndrome''. The identity of $e$ is uniquely determined by this ``syndrome'',
since the measurement process projects any linear combination of errors to
an error in the Pauli group.

\subsection{CSS-Stabilizer Errors}

In this case the stabilizer generators contain either products only of $Z$'s
(``$Z$-type'')\ or products only of $X$'s (``$X$-type''). Suppose we wish to
measure a $Z$-type stabilizer generator. The $+1$ eigenstates of such a
generator are the ``even parity states'', i.e., those states containing an
even number of $|1\rangle $'s. Prepare an ancilla in the encoded $
|0_{L}\rangle $ state (below we discuss how). Then for each data qubit where
the given stabilizer generator has a $Z$ (not an $I$) apply a controlled-$
\overline{X}$ from this qubit to the ancilla. The ancilla will flip every time
the data qubit was a $|1\rangle $, so measuring the ancilla at the end and
finding it in $|0_{L}\rangle $ will indicate no error (even number of
flips), whereas $|1_{L}\rangle $ will indicate an error (odd number of
flips). Distinguishing between $|0_{L}\rangle $ and $|1_{L}\rangle $ amounts
to measuring $\overline{Z}$ on the ancilla, which we can do directly by
measuring $Z$ on all those ancilla qubits whose $\overline{Z}$ has a $Z$.

Now suppose we wish to measure an $X$-type stabilizer element. The same
procedure as for $Z$-type generators can be applied, with one modification:$
\;$a Hadamard transform 
\begin{equation}
R=\frac{1}{\sqrt{2}}\left( 
\begin{array}{cc}
1 & 1 \\ 
1 & -1
\end{array}
\right)
\end{equation}
must be applied before and after the controlled-$\overline{X}$ operation. The
effect of the Hadamard transform before the controlled-$\overline{X}$ operation
is to change the corresponding qubit into the $Z$-eigenbasis, whence the $Z$
-type construction applies. The second Hadamard transform returns the qubit
to the original basis. This construction is shown schematically in Fig.~\ref
{fig:measurement}.

Note that since $\overline{X}$ is in the normalizer, it commutes with all
stabilizer errors. This means that any such error occuring on the ancilla
before the $\overline{X}$ is equivalent to the same error after the $\overline{X}$,
and therefore the error has no effect. In other words, neither does the
ancilla ever leave the DFS under the application of $\overline{X}$, nor can an
error on the ancilla propagate back to the data qubits.\footnote{
This DFS-construction is different than in the usual QECC-stabilizer
construction, where multiple control operations to the same ancilla-qubit
are not fault-tolerant because they are not transversal. There multiple {\sc 
CNOT}'s from different data qubits to the same ancilla qubit can cause
errors to spread catastrophically if the ancilla qubit undergoes a phase
error (recall that under {\sc CNOT}, $I\otimes Z\mapsto Z\otimes Z$).} Note
further that since the ancilla is at all times unentangled from the data
qubits the measurement is non-destructive on the data qubits.

What if a stabilizer error occurs on the data qubits right after the
application of the Hadamard gate? This can clearly present a problem, since
it may for example flip the data qubit controlling the $\overline{X}$ applied to
the ancilla. One (standard) way of dealing with such errors is to repeat the
measurement several times in order to improve our confidence in the result.
An alternative is to use concatenated codes \cite
{Knill:97a,Aharonov:96,Zalka:96,Knill:98}. This will be of use if the
stabilizer error is correctable by the transformed code, i.e., if we can
verify that the conditions of Theorem 2 are satisfied. Then we can use the
DFS at the lowest level, and concatenate it with the QECC it transforms into
under the stabilizer errors (see Ref. \cite{Lidar:PRL99} for concatenated
DFS-QECC in the collective decoherence model). Now, recall the CSS form of
the normalizer elements, Eq.~(\ref{eq:ZbarCSS}). For every Hadamard
transform in the first set (i.e., before the controlled-$\overline{X}$
operations) on a qubit in a position corresponding to an $X$ in an $X$-type
stabilizer generator, the normalizer elements transform by having $X$ and $Z$
interchange in this position. In the standard form of Eq.~(\ref{eq:ZbarCSS}), if this happens to be the first qubit then $\overline{Z}\longmapsto
X\otimes M_{Z}\otimes I$, which anticommutes with the original $\overline{Z}$, and $\overline{X}\longmapsto Z\otimes I\otimes M_{X}$, which in turn
anticommutes with the original $\overline{X}$. If the position of the $X$ in
the $X$-type stabilizer generator is where $M_{Z}$ has a $Z$, then $
\overline{Z}\longmapsto Z\otimes M_{Z}^{\prime }\otimes I$, where $
M_{Z}^{\prime }$ has that $Z$ changed into an $X$. This transformed $
\overline{Z}$ anticommutes with the original $\overline{X}$. Similarly, $
\overline{X}\longmapsto X\otimes I\otimes M_{X}^{\prime }$ with an $X$
changed into a $Z$, and this transformed $\overline{X}$ anticommutes with
the original $\overline{Z}$. Thus the conditions of Theorem 2 are again
satisfied.

The second set of Hadamard transforms restores the original normalizer. One
then proceeds to measure the next stabilizer generator. We thus see that
this measurement procedure is fault-tolerant of stabilizer errors.

\subsection{General Stabilizer Errors}

In the non-CSS case the stabilizer generators may contain $Y$'s as well, so
our analysis above requires some modifications. The unitary operation that
transforms $Y$ to $Z$ is 
\begin{equation}
Q=\frac{1}{\sqrt{2}}\left( 
\begin{array}{cc}
1 & -i \\ 
1 & i
\end{array}
\right) .
\end{equation}
It also maps $Z\mapsto X\mapsto Y$. When this operation is applied
immediately before the controlled-$\overline{X}$ to the ancilla and
immediately after it for every $Y\;$in the stabilizer, the $Z$-type
construction applies again. However, for the purpose of concatenation we
need to check that the procedure is still fault-tolerant of stabilizer
errors. The normalizer generators now have the form of Eqs.~(\ref{eq:Zbarj}),(\ref{eq:Xbarj}). Every time a Hadamard or $Q$ operation is applied, $
Z\mapsto X$ in a single position in $\overline{Z}$. Similarly, $Z\mapsto X$,
or $X\mapsto Z$ (if Hadamard) or $Y$ (if $Q$) in a single position in $
\overline{X}$.

The case of the transformed $\overline{Z}$ is trivial: if $Z\mapsto X$
anywhere then the transformed $\overline{Z}$ anticommutes with the original $
\overline{Z}$. Consider the transformed $\overline{X}$. The possibilities
are:\ (i) $X_{1}\mapsto Z_{1}$ or $Y_{1}$, (ii) $Z\mapsto X$ in the $N_{Z}$
part, (iii) $X\mapsto Z$ or $Y$ in the $M_{X}$ part. In all these cases it
is easily verified that the transformed $\overline{X}$ anticommutes with the
original $\overline{X}$. Therefore the measurement procedure is
fault-tolerant also in the non-CSS case.

\section{Outlook: Implications for the Independent-Errors Model}

\label{Outlook}

The methods we have introduced in this paper need not be restricted to
stabilizer-errors. In this section we briefly touch upon the implications of
our construction for universal quantum computation in the independent errors
model, when stabilizer-errors are taken into account as well. We thus
generalize the standard treatment of stabilizer codes \cite{Gottesman:97a},
where stabilizer errors that may occur during the course of gate
implementation are ignored. However, we are here only able to consider
independent single-qubit errors, so that the inclusion of the special type
of correlated many-body errors represented by the stabilizer-errors is a
rather unrealistic error model. The main importance of the result presented
here is that it suggests an alternative route to universal quantum
computation that is fault tolerant with respect to error {\em detection},
and is highly parallelizable. We believe that this may lead to an improved
threshold for fault tolerant computation in the setting of concatenated
codes \cite{Knill:98}.

Let us recall the error detection and correction criteria for a stabilizer
code $Q=\{q_{k}\}$ to be able to deal with all single qubit errors: 
\begin{equation}
\forall i,j,\alpha ,\beta \text{ }\exists k\text{ s.t. }\{q_{k},\sigma
_{i}^{\alpha }\otimes \sigma _{j}^{\beta }\}=0  \label{eq:12QECC-cond}
\end{equation}
Can we implement encoded $SU(2)$ operations in the independent errors model
similarly to what we did above for stabilizer-errors? To do so we need to
make sure that the errors do not become part of the sequence of transformed
normalizers. The important difference compared to the stabilizer-errors case
is that now the errors are ``small'' (single-body), which means that we must
avoid using a single-qubit Hamiltonian as a central gate (for it is a
normalizer element which will not be distinguishable from an error). If we
restrict ourselves to using two-body Hamiltonians as central gates (which we
can always do -- recall the \ comment at the end of Section~\ref{generalXZ}
), then we run into a similar problem regarding the two-body form of Eq.~(\ref{eq:12QECC-cond}), i.e., if the central gate uses the Hamiltonian $
\sigma _{i}^{\alpha }\otimes \sigma _{j}^{\beta }$ then we will not be able
to correct the two errors $\sigma _{i}^{\alpha }$ and $\sigma _{j}^{\beta }$. However, as we now show, as long as we use a two-body central gate it is
nearly always possible to satisfy the error {\em detection} criterion, $
\forall i,\alpha $ $\exists k$ s.t. $\{q_{k},\sigma _{i}^{\alpha }\}=0$.

Let us demonstrate this explicitly for Steane's 7-qubit code \cite
{Steane:96a}. This is a CSS code encoding one qubit into seven, and in
standard form has the normalizer: 
\begin{eqnarray}
\overline{X} &=&X_{1}X_{5}X_{6}  \nonumber \\
\overline{Z} &=&Z_{1}Z_{3}Z_{4}.
\end{eqnarray}
Consider the gate construction [derived from Eq.~(\ref{eq:An})] 
\begin{equation}
\exp (i\theta \overline{Z})=T_{X_{1}Z_{3}}\circ \exp (i\theta Y_{1}Z_{4}).
\end{equation}
The normalizer transforms as:

\begin{eqnarray}
\overline{X}&\stackrel{X_{1}Z_{3}}{\longmapsto }&\overline{X}\stackrel{
Y_{1}Z_{4}}{\longmapsto }\cos (2\theta )\overline{X}+i\sin (2\theta)Z_{1}Z_{4}X_{5}X_{6}\stackrel{X_{1}Z_{3}}{\longmapsto }\cos (2\theta )
\overline{X}+\sin (2\theta )\overline{Y} =\overline{X}\exp (2i\theta 
\overline{Z})  \nonumber \\
 \overline{Z}&\stackrel{X_{1}Z_{3}}{\longmapsto }&Y_{1}Z_{4}\stackrel{
Y_{1}Z_{4}}{\longmapsto }Y_{1}Z_{4}\stackrel{X_{1}Z_{3}}{\longmapsto }
\overline{Z}.
\end{eqnarray}
We see that at no point does a single-qubit error become part of the
transformed normalizer, so that all single qubit errors are detectable. On
the other hand, while we can always detect the occurrence of both the $Y_{1}$
and $Z_{4}$ errors, we cannot distinguish between them after the first gate
has been applied (since our normalizer is $Y_{1}Z_{4}$ at that point). Since
we might accidentally try to reverse the error $Y_{1}$ when in fact the
error $Z_{4}$ has taken place, this means that our construction is fault
tolerant only for error detection. Similarly, the gate construction 
\begin{equation}
\exp (i\theta \overline{X})=T_{Z_{1}X_{5}}\circ \exp (i\theta Y_{1}X_{6})
\end{equation}
yields 
\begin{eqnarray}
 \overline{X}&\stackrel{Z_{1}X_{5}}{\longmapsto }&Y_{1}X_{6}\stackrel{
Y_{1}X_{6}}{\longmapsto }Y_{1}X_{6}\stackrel{Z_{1}X_{5}}{\longmapsto }
\overline{X}  \nonumber \\
\overline{Z}&\stackrel{Z_{1}X_{5}}{\longmapsto }&\overline{Z}\stackrel{
Y_{1}X_{6}}{\longmapsto }\cos (2\theta )\overline{Z}+i\sin (2\theta
)X_{1}Z_{3}Z_{4}X_{6}\stackrel{Z_{1}X_{5}}{\longmapsto }\cos (2\theta )
\overline{Z}+\sin (2\theta )\overline{Y}=\overline{Z}\exp (-2i\theta 
\overline{X}).
\end{eqnarray}
which also satisfies the error detection (but not correction) condition for
single-qubit errors, in that no single-qubit error becomes part of the
transformed stabilizer.

Let us now consider the general stabilizer case. Recall once more the
standard form of the normalizer, Eqs.~(\ref{eq:Zbarj}),(\ref{eq:Xbarj}). Our
gate construction acts by transforming one of the normalizer elements to
two-body form, where it is applied as the central $\theta $-gate, and then
is transformed back to its standard form. All other normalizer elements are
left unchanged until the application of the central gate, with which they
anticommute. At this point each $\overline{Z}$ [$\overline{X}$] is
multiplied by $\exp (-2i\theta \overline{X})$ [$\exp (2i\theta \overline{Z})$]. The final sequence of gates flips these normalizer elements back and
forth between $\exp (-2i\theta \overline{X})$ and $\exp (-2i\theta \overline{
Y})$ [$\exp (2i\theta \overline{Z})$ and $\exp (2i\theta \overline{Y})$] (recall the analysis in Section~\ref{CSS1qubit}). All these operations have
the effect of expanding, rather than shrinking the normalizer elements, as
seen in the example of the 7-qubit code above.

The ability to error-detect at each point thus translates to the question of
whether any normalizer element ever becomes a single-body Hamiltonian under
this sequence of transformations. It is not hard to see from the above
description of the orbit of the normalizer that this can only be the case if
in the standard form the normalizer contains a single-body element to begin
with. This is certainly possible, as indeed shown in our $Q_{2X}$ example
(Section~\ref{Q2X}), where $\overline{X}=XIII$. However, it is not the case
for most interesting stabilizer codes, i.e., those offering protection
against arbitrary single-qubit errors. Such codes must have ``large''
normalizer elements since they may not contain any single-qubit errors to
begin with. We conclude that our $\overline{SU(2)}$ construction using just
two-qubit Hamiltonians works for all stabilizer codes of interest, in the
sense that it is fault-tolerant with respect to error detection.

To complete the repertoir of universal operations the $\overline{\text{{\sc 
CNOT}}}$ gate is still needed. The discussion given in Section~\ref{CNOT}
applies here as well, with the modification that for non-CSS stabilizer
codes it is once again necessary to apply two-body central gates. Fault
tolerant measurement of the error syndrome can be done using the standard
techniques available for stabilizer codes \cite{Gottesman:97a}.

\section{Summary and Conclusions}
\label{Summary}

In a previous paper \cite{Lidar:00a} we derived conditions for the
existence of class of decoherence-free subspaces (DFSs) defined by having
Abelian stabilizers over the Pauli group. In this sequel paper we addressed
the problem of universal, fault-tolerant quantum computation on this class
of DFSs. The errors in this model are the elements of the stabilizer, and
thus are necessarily correlated. This model is complementary to the standard
model of quantum computation using stabilizer quantum error correcting codes
(QECCs), where the errors that are correctable by the code anticommute with
the stabilizer (rather than being part of it). The correlation between
errors in the present model implies no spatial symmetry in the system-bath
interaction, unlike in most previous studies of computation on DFSs (which
considered the ``collective decoherence'' model, and where the stabilizer
is non-Abelian). Therefore our present results significantly enlarge the
scope of the theory of DFSs.

It turns out that even though the class of DFSs we considered are
Pauli-group stabilizer
codes, the usual universality constructions do not apply, because of the
different error-model we assume. Our alternative construction of a set of
universal quantum gates resorts to the early ideas about universal quantum
computation, except that our operations all act on {\em encoded} (DFS)
qubits: we showed how to implement arbitrary single-encoded-qubit
operations [the $\overline{SU(2)}$ group] and $\overline{\text{{\sc CNOT}}}$
gates between pairs of encoded qubits. The challenge here was to show how
to accomplish this implementation using only physically reasonable
Hamiltonians, i.e., those involving no more than two-body interactions. To
do so, we switched from the usual point of view treating the normalizer
elements (i.e., the operations that preserve the DFS) as gates, to one
where these elements are considered as many-body Hamiltonians. We then
introduced a procedure whereby these Hamiltonians could be simulated using
at most two-body interactions. The gate sequence implementing this
simulation does not preserve the DFS except at the beginning and end.
Throughout the execution of the gates the DFS states are exposed to the
stabilizer-errors. However, we showed that in fact the DFS is transformed
into a sequence of stabilizer codes, each of which is capable of detecting
and correcting the original stabilizer-errors. Moreover, we showed that
these errors can be diagnosed in a fault-tolerant manner, i.e., without
introducing new errors as a result of the associated measurements. In all,
we showed how by using this type of hybrid DFS-QECC approach, universal,
fault-tolerant quantum computation can be implemented.

Our results have implications beyond computation on DFSs. We briefly
considered here also the question of whether our techniques can be used to
compute fault-tolerantly in the standard stabilizer error-model. We found
the answer to be affirmative for the purpose of single-qubit error
detection, but not correction. While this is interesting in its own right
because of the new universality construction we introduced, it may also have
important implications for the question of quantum computation using
concatenated codes. The reason is that our construction is highly
parallelizable, meaning that it requires a very small number of operations
during which the encoded information is exposed to errors. We speculate that
this can significantly reduce the threshold for fault-tolerant quantum
computation.

Finally, an interesting open question is whether the methods developed
here are applicable to the problem of universal quantum
computation on other classes of DFSs.

\section{Acknowledgments}

This material is based upon work supported by the U.S. Army Research Office
under contract/grant number DAAG55-98-1-0371, and in part by NSF
CHE-9616615. We would like to thank Dr. Daniel Gottesman for very useful
correspondence.

\appendix

\section{Why the 4-Block Implementation of $\overline{\text{{\sc CNOT}}}$ is
not Fault-Tolerant for non-CSS Stabilizers}

\label{app:A}

The construction of the $\overline{\text{{\sc CNOT}}}$ in Ref. \cite
{Gottesman:97a} uses a series of bitwise {\sc CNOT}'s (along with some other
operations) acting between pairs of qubits in $4$ different blocks. Let us
calculate the result of applying bitwise {\sc CNOT}'s on $I^{\otimes
K}\otimes \overline{X}$ (i.e., on two out of the four blocks). Recall that for a
non-CSS code $\overline{X}=X\otimes N_{Z}\otimes M_{X}$ [Eq.~(\ref{eq:Xbarj})].
Therefore it follows from Eq.~(\ref{eq:transf-table}) that 
\begin{equation}
I^{\otimes K}\otimes \overline{X}\longmapsto \left[ I\otimes N_{Z}\otimes
I^{\otimes K-1-r}\right] \otimes \overline{X},
\end{equation}
i.e., the $Z$'s are copied backwards into the first block. Therefore the
normalizer on the first block now contains $I\otimes N_{Z}\otimes I^{\otimes
K-1-r}$. This element obviously commutes with both the original $\overline{X}$
and $\overline{Z}$ [Eq.~(\ref{eq:Zbarj})], but does not equal either. Therefore
it must be in the original stabilizer $Q$. Turning this around, we see that
an error $e\in Q$ has become part of the new normalizer $N_{j}(Q_{j})/Q_{j}$
which is catastrophic since this error is now undetectable.


\newpage

\begin{figure}[!htb]
\hspace{0.2\textwidth}
\caption{Fault-tolerant circuit implementing $\exp (i\protect\theta 
\overline{Z})$ for the $Q_{2X}$ subgroup. The transformed $\overline{Z}$ is
shown at each gate, and directly below the original normalizer element
that it anticommutes with.}
\label{fig:gates-Q2X}
\end{figure}

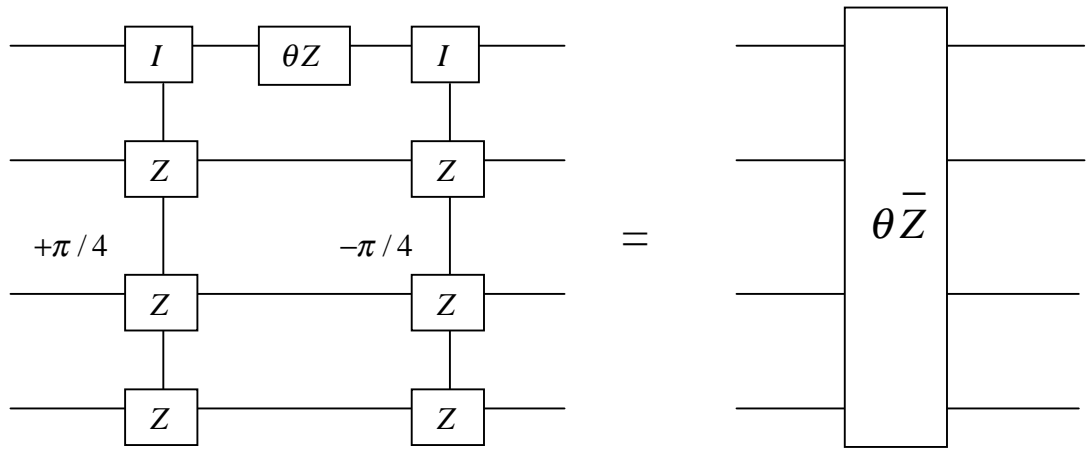
\begin{figure}[!htb]
\hspace{0.2\textwidth}
\caption{Parallel implementation of $\protect\theta 
\overline{Z}$ for the $Q_{2X}$ subgroup.}
\label{fig:parallel}
\end{figure}

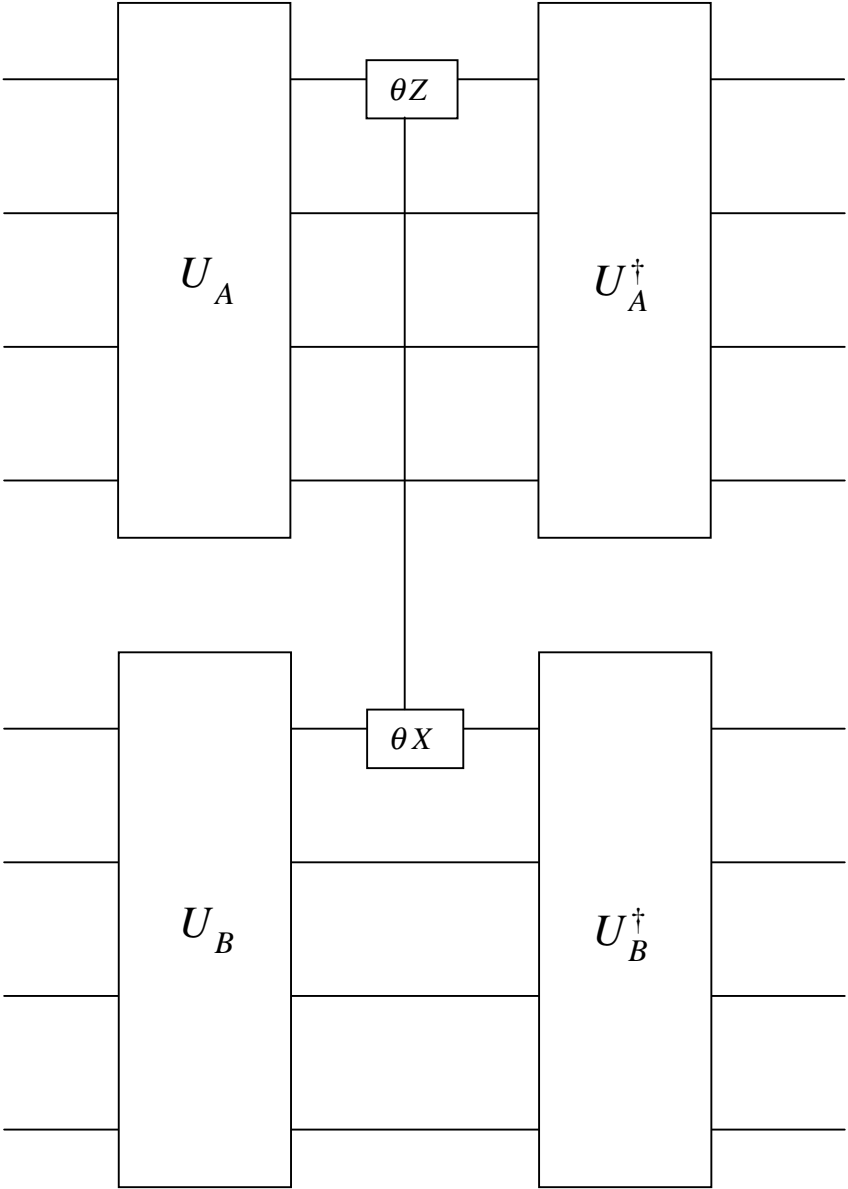
\begin{figure}[!htb]
\hspace{0.2\textwidth}
\caption{Fault-tolerant implementation of $\exp (i\protect\theta 
\overline{Z}\overline{X})$ needed to generate ${\sc CNOT}$.}
\label{fig:CNOT}
\end{figure}

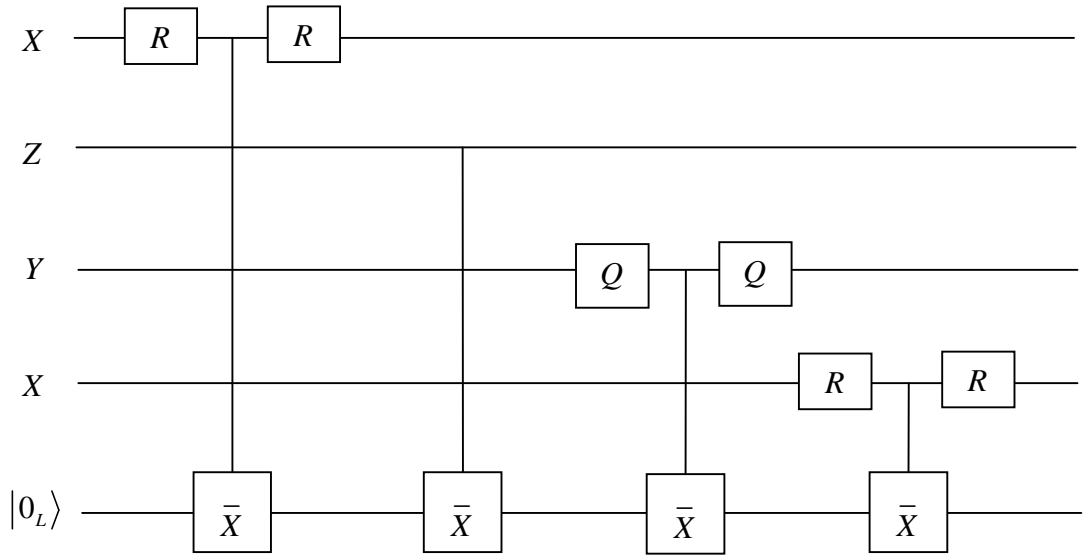
\begin{figure}[!htb]
\hspace{0.2\textwidth}
\caption{Measurement of the stabilizer element $XZYX$.}
\label{fig:measurement}
\end{figure}

\end{document}